\documentclass{article}%
\usepackage{amsmath}
\usepackage{amsfonts}
\usepackage{amssymb}
\usepackage{graphicx}%
\newtheorem{theorem}{Theorem}

\newtheorem{definition}[theorem]{Definition}

\newtheorem{notation}[theorem]{Notation}

\newtheorem{proposition}[theorem]{Proposition}

\newtheorem{summary}[theorem]{Summary}
\newenvironment{proof}[1][Proof]{\noindent\textbf{#1.} }{\ \rule{0.5em}{0.5em}}
\begin{document}

\title{Relating Spin Foams and Canonical Quantum Gravity: $(n-1)+1$ formulation of
$nD$ spin foams.}
\author{Suresh K\ Maran\\Department of Physics and Astronomy\\University of Pittsburgh\\Pittsburgh PA-15260}
\maketitle

\begin{abstract}
In this article we lay foundations for a formal relationship of spin foam
models of gravity and BF theory to their continuum canonical formulations.
First the derivation of the spin foam model of the BF\ theory from the
discrete BF theory action in $n$ dimensions is reviewed briefly. By foliating
the underlying $n$ dimensional simplicial manifold using $n-1$ dimensional
simplicial hypersurfaces, the spin foam model is reformulated. Then it is
shown that spin network functionals arise naturally on the foliations. The
graphs of these spin network functionals are dual to the triangulations of the
foliating hypersurfaces. Quantum Transition amplitudes are defined. I
calculate the transition amplitudes related to $2D$ BF theory explicitly and
show that these amplitudes are triangulation independent. The application to
the spin foam models of gravity is discussed briefly.

\end{abstract}

\section{Introduction}

During the late part of the last decade, there has been a vigorous activity in
the area of combinatorial quantization of \ the theories such as BF
theory\footnote{A BF theory in $n$ dimensions and for a group $G$ refers to
field theory defined by the action $S=\int B\wedge F$ . Here $B$ is a $n-2$
form which takes values in dual Lie algebra of $G$ and $F$ is a 2-form is the
cartan curvature of a $G$-connection $A$. The free variables of the theory are
$B$ and $A$.} \cite{BF} and gravity, generally referred to as the spin foam
quantization. The general notion of a spin foam model was motivated by at
least three examples: the Regge-Ponzano model, which is a construction of
simplicial quantum geometries using $6J$ symbols of the group $SU(2)$
\cite{Rg1}; the abstract spin networks of Roger Penrose, who derives spatial
structures from the interchange of angular momentum \cite{RP} and the
evolutions of the Rovelli-Smolin spin network functionals, which are the
kinematical quantum states of canonical quantum gravity \cite{R5}. Casual
evolution and dual formulation of spin foams were proposed by Fotini
Markopoulou \cite{fms}, \cite{fm} and Lee Smolin \cite{fms}. I refer to Baez
\cite{bz1} for a nice introduction to spin foam models and we refer to Perez
\cite{AP2} for an up-to-date review of the spin foam models and a
comprehensive set of references.

The concept of spin foam is very general and there are various specific spin
foam models that are available in the research literature \cite{AP2}. A spin
foam model of the four dimensional $SO(4)$ BF theory called the Ooguri model
\cite{oo2} can be derived directly from its discretized action. From this
model a spin foam model of Riemannian gravity can be derived by imposing a set
of constraints called the Barrett-Crane constraints \cite{bc1}. \emph{Here by
`the spin foam models' we specifically refer to these models of BF theory,
gravity and their variations.}

One of the interesting problems in quantum gravity is how to relate the spin
foam models of gravity to its canonical formulation of gravity \cite{As},
\cite{RS}, \cite{TT1}, \cite{R3}. Here we take the point of view of seeing how
close we can bring a spin foam model to it's canonical quantum formulation,
instead of assuming the existence of a precise relation between them.
Canonical quantum gravity is formulated on continuum manifolds, while the spin
foams are formulated on simplicial manifolds\ (or on 2-complexes \cite{bz1}).
In general the canonical formulation requires the underlying $n$ dimensional
manifold of the theory to be expressible in an $(n-1)+1$ form. In the same
spirit, here we foliate the $n$ dimensional simplicial manifold. The foliation
is made up of a one parameter family of simplicial $(n-1)$ dimensional
hypersurfaces\footnote{We restrict ourselves to manifolds that are foliatable
by a one parameter family of simplicial hypersurfaces.}. Between any two
consecutive hypersurfaces we have a one-simplex thick slice of the simplicial manifold.

To each edge ($(n-1)$-simplex) of the simplicial manifold is associated a
parallel propagator $g$ which plays the role of a discrete connection. To make
a parallel to the canonical quantization we make an important identification.
I find that the parallel propagators associated with the edges in the
foliating hypersurfaces can be thought of as the analog of the continuum
connection in the (coordinate) time direction. In the canonical quantization,
the field equation corresponding to this component of the connection is the
Gauss constraint \cite{As}, which on quantization leads to spin network
functionals \cite{RS}. Remarkably the same idea works in the spin foam
quantization obtained by the path integral quantization of the theory defined
by the discretized BF action $S$. It just happens that the integration of the
Feynman weight $e^{iS}$ with respect to the parallel propagators associated
with the edges of the hypersurfaces results in a product of spin network
functionals. These spin network functionals are defined on the parallel
propagators associated with the edges that go between the hypersurfaces and
the graphs that are dual to the triangulation of the foliating hypersurfaces.
All our work is built around this observation of the appearance of spin
network functionals. Since the spin foam model of gravity is obtained by
imposing the Barrett-Crane constraints on the BF\ spin foam model, we believe
we can carry over this result to gravity.

Each of the one-simplex thick slices of the simplicial manifold can be
considered to define a discrete coordinate time instant. The set of parallel
propagators which are associated with the edges that go between two
consecutive foliating hypersurface can be considered to contain the physical
(connection) information of the theory at a particular discrete time for a
given triangulation. Spin network states can be defined as functions of these
discrete connections. Using the path integral formulation, we define a spin
network state to spin network state elementary transition amplitude matrix.

This article has been made as self-contained as possible. In this article we
first focus on BF theory for an arbitrary compact group and discuss gravity
afterwards. In section two we review the derivation of BF spin foam model.

In section three we discuss how the partition function of the BF\ theory can
be expressed in terms of the spin network functionals that are obtained by
integrating the Feynman weight $e^{iS}$ with respect to the parallel
propagators associated with the edges in the foliating simplicial
hypersurfaces. In section four we discuss the details of these spin network
functionals. I show that these spin network functionals are orthonormal in the
obvious inner product.

In section five we discuss the elementary transition amplitudes using the path
integral formulation. I discuss this in the form of a connection formulation
and spin network formulation.

In section six we discuss two dimensional BF theory. I explicitly calculate
the elementary transition amplitudes. I find that the transition matrix is
symmetric, non-unitary and is independent of triangulation.

In section seven we discuss $2+1$ BF theory ($2+1$ Riemannian) gravity very briefly.

In section five we define the elementary transition amplitude matrix for
gravity by including the Barrett-Crane constraints in the definition of
BF\ elementary transition amplitude matrix. This is similar to that of
Reisenberger \cite{rz4} defined in an unfoliated context. In section six, we
observe that, in the case of Lorentzian Barrett-Crane model, in the asymptotic
limit, the foliating hypersurfaces behave as spatial hypersurfaces.

\section{Review of the spin foam derivation}

The review in this section follows that of Baez \cite{bz1}. Advanced readers
may skip or quickly glance through this section. The term `edge integral' is
introduced in this section and is used widely in this article.

Consider an $n$ dimensional manifold $M$ and a $G$-connection $A$, where $G$
is a compact linear group. Let $F$ be a curvature 2-form of the connection
$A$. Also let $B$\ be a dual Lie algebra valued $n-2$ form. Then the continuum
BF theory is defined by the following action:%
\begin{equation}
S_{c}=\int_{M}Tr(B\wedge F). \label{eq.1}%
\end{equation}

The spin foam model for this action is derived by calculating the partition
function corresponding to the discretized version of this action \cite{bz1},
\cite{oo2}, \cite{Lfr1}. Let the manifold be triangulated by a simplicial
lattice. Each $n$-simplex $s$ is bounded by $n+1$ $(n-1)$-simplices called the
edges $e$ of $s$. In turn each $(n-1)$-simplex is bounded by $n$
$(n-2)$-simplices called the bones.

To discretize the BF action, associate a group element $g_{e}$ with each edge
$e$ of the lattice.%
\begin{figure}
[ptbh]
\begin{center}
\includegraphics[
height=2.9084in,
width=3.5189in
]%
{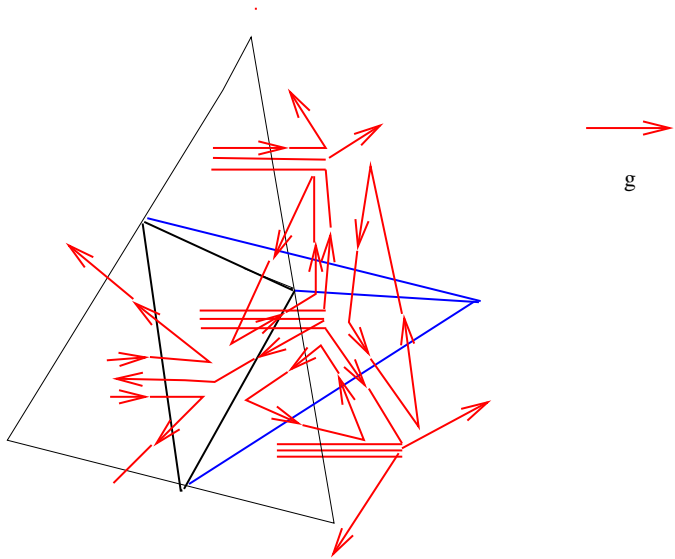}%
\caption{Holonomies.}%
\label{fig.1}%
\end{center}
\end{figure}
This is considered to be the parallel propagator of the connection $A$ related
to moving a $G$-vector from a given point in one of the $n$-simplices to an
adjacent one through the edge $e$. Then the discrete analog of the curvature
$F$ is $\ln H_{b}$, where $H_{b}={\prod_{e\supset b}}g_{e}$ is the holonomy
around each bone\footnote{There is an arbitrariness in the base point of the
holonomy, but it will not be a problem as we will see soon.} (figure
(\ref{fig.1})) and $\ln$ is a map from the group space to its Lie algebra
space. Then the discrete BF action is%
\begin{equation}
S_{d}=\sum_{b}Tr(B_{b}\ln H_{b})
\end{equation}
Here $B_{b}$ =$\int_{b}B$ is the discrete analog of $B$ and the trace is taken
in the Lie algebra index. Then the\ quantum partition function is calculated
using the path integral formulation\footnote{Please notice that the action is
real because the Lie algebra co-ordinates are real for compact groups.}:%
\begin{align}
Z  &  =\int\prod_{e}dg_{e}\prod_{b}dB_{b}\exp(iS_{d})\nonumber\\
&  =\int\prod_{e}dg_{e}\prod_{b}dB_{b}\exp(i\sum_{b}Tr(B_{b}\ln H_{b})).
\end{align}

Integration is over each group variable $g_{e}$ and over each Lie algebra
valued $B_{b}$ variable of the triangulated manifold.\ Here $dg_{e}$ is the
Haar measure on the group. Doing the integration over the $B_{b}$ variables
results in the following:%
\begin{equation}
Z=\int\prod_{e}dg_{e}\prod_{b}\delta(H_{b}), \label{eq.3}%
\end{equation}
where $\delta(H)$ is the delta functional on the group. Since the group is
compact, the expansion of the delta functional is given by \cite{de1}%
\begin{equation}
\delta(H)=\sum_{J}d_{J}Tr(\rho_{J}(H)),
\end{equation}
where $\rho_{J}(H)$ is the $J$ representation of the group (tensor indices not
shown) and $d_{J}$ is the dimension of the representation. Substituting this
into equation $\left(  \ref{eq.3}\right)  $ we get%
\begin{equation}
Z=\int\prod_{e}dg_{e}\prod_{b}\sum_{J_{b}}d_{J_{b}}Tr(\rho_{J_{b}}%
(\prod_{e\supset b}g_{e})) \label{eq.4.0}%
\end{equation}%
\begin{equation}
=\sum_{\left\{  J_{b}\right\}  }\left[  \left(  \prod_{b}d_{J_{b}}\right)
\left(  \int\prod_{b}dg_{e}Tr(\prod_{e\supset b}\rho_{J_{b}}(g_{e}))\right)
\right]  \label{eq.4h}%
\end{equation}%
\begin{equation}
=\sum_{\left\{  J_{b}\right\}  }\left[  \left(  \prod_{b}d_{J_{b}}\right)
\left(  \int\prod_{b}Tr\bigotimes_{b\subset e}\rho_{J_{b}}(g_{e})\prod
_{e}dg_{e}\right)  \right]  , \label{eq.4}%
\end{equation}
where $Tr$ denotes the required summing operations from the trace operations
in the previous line. This equation will be used in the next section to make
an $(n-1)+1$ splitting of the theory. The integrand of the quantity in the
second parentheses is the $g_{e}$ integration of the tensor product of the
representation matrices $\rho_{J_{b}}(g_{e})$ that were part of the holonomy
around the $n$ bones of the edge $e$. This quantity can be rewritten as a
product of orthonormal basis of intertwiners $i$ as follows:%
\begin{equation}
\int dg\bigotimes_{b\supset e}\rho_{J_{b}}(g)=\sum_{i_{e}}i_{e}\bar{\imath
}_{e}. \label{eq.5.0}%
\end{equation}
The integral on the left hand side of this equation will be referred to as an
\textbf{edge integral}. The bar denotes adjoint operation.\emph{ Each one of
the two intertwiners corresponds to one of the two sides of an edge of a
simplex. }Please refer to the appendices for more information about the edge integrals.

\textit{The mathematical fact that the edge integral splits into two
intertwiners is a critical reason for the emergence of the spin foam models
from the path integral formulation of the discretized BF theory. }Each of the
intertwiners is associated with one of two sides of the edge\textit{.}When
this edge integral formula is used in equation $\left(  \ref{eq.4}\right)  $
and all the required summations are performed, it is seen that each index of
each intertwiner corresponding to an inner side of an edge of each simplex
only sums with an index of an intertwiner corresponding to an inner side of
another edge of the same simplex. Because of this the partition function $Z$
splits into a product of terms, with each term interpreted as a quantum
amplitude associated with a simplex in the triangulation.

Finally the formula for the partition function in $n$ dimensions is given by%
\begin{equation}
Z=\sum_{\left\{  J_{b},i_{e}\right\}  }\left(  \prod_{b}d_{J_{b}}\right)
\prod_{s}Z(s), \label{eq.6}%
\end{equation}
where $Z(s)$ is the quantum amplitude associated with the $n$-simplex $s$ and
$d_{J_{b}}$ is interpreted as the quantum amplitude associated with the bone
$b$. This partition function may not be finite in general. The set $\left\{
\boldsymbol{J}_{b,}\boldsymbol{i}_{e}\right\}  $ of all $J_{b}$'s and $i_{e}%
$'s is called a \textbf{coloring} of the bones and the edges.

\section{The $(n-1)+1$ splitting of the n dimensional BF spin foam
models.\label{sec.2}}

Consider a smooth $n$ dimensional manifold $M$ triangulated by a simplicial
lattice. I assume that the following properties hold for the
triangulation\footnote{Please note that these conditions restrict the set of
all allowable topologies for $M.$ But I believe we will be able include the
excluded topologies by adding additional constructs to our formulation.}%
\footnote{It appears that this technique works even if there is a topology
change at a hypersurface. Please see the discussion on the $1+1$ formulation
in section 4.}:

\begin{enumerate}
\item The simplicial manifold can be foliated by a discrete one parameter
family of $n-1$ dimensional simplicial hypersurfaces made of the edges of the triangulation,

\item The foliation is such that there are no vertices of the lattice in
between the hypersurfaces of the family,

\item The hypersurfaces do not intersect or touch each other at any point, and

\item The slice of the manifold in between any two consecutive hypersurfaces
is always one-simplex thick.
\end{enumerate}

\begin{figure}
[ptbh]
\begin{center}
\includegraphics[
height=4.8732in,
width=4.9329in
]%
{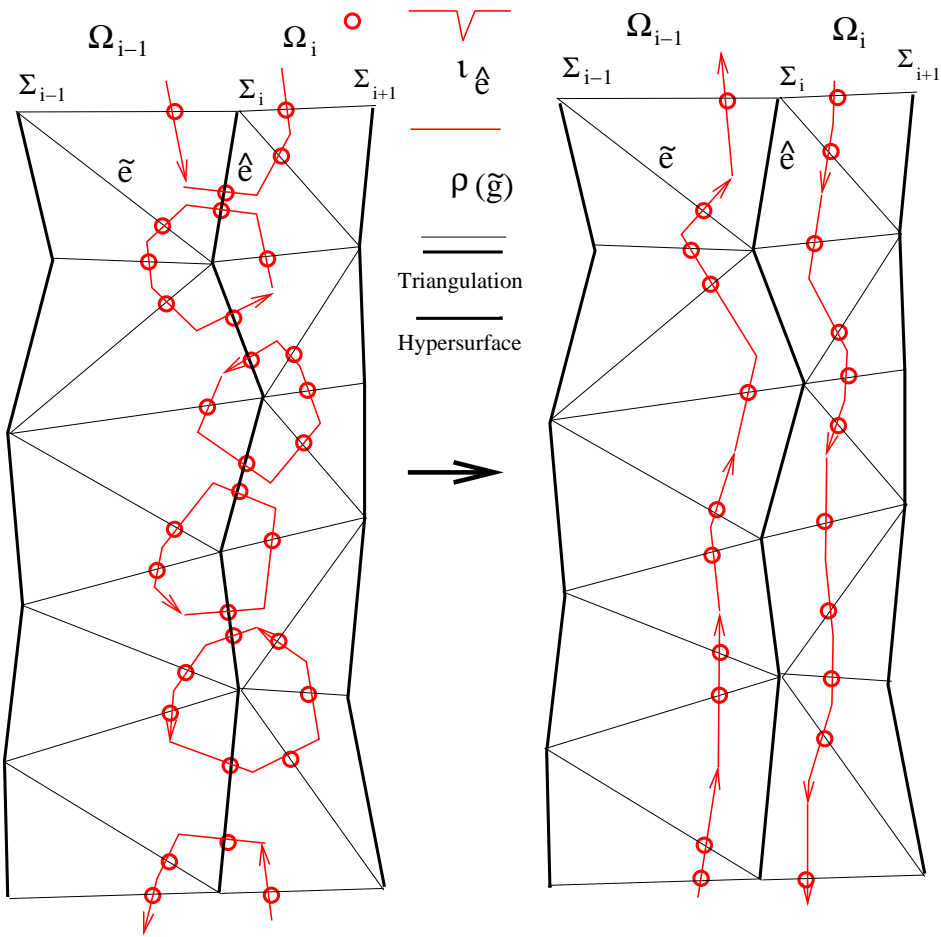}%
\caption{Before and after integration with respect to $dg_{\hat{e}}$, $\hat
{e}$ $\in$ $\Sigma_{i}.$}%
\label{fig.spinder}%
\end{center}
\end{figure}

Now let's define the following notations. Please see figure (\ref{fig.spinder}).

\begin{notation}
Let $\left\{  \Sigma_{i}\right\}  $ be a sequence of simplicial hypersurfaces,
ordered by an integer $i$, which is a foliation of the triangulation of $M$
such that the above properties hold.
\end{notation}

\begin{notation}
Let $\Omega_{i}$ be the piece of the simplicial manifold $M$ between
$\Sigma_{i}$ and $\Sigma_{i+1}$. This $\Omega_{i}$ has the thickness of a one-simplex.
\end{notation}

Now there are two types of edges and bones in the lattice, those that which
lie on the hypersurfaces and those that go between the hypersurfaces.

\begin{notation}
Let the edges which lie on the hypersurfaces be represented with a
\textbf{cap} on them, as in $\hat{e}$, and those that go between the foliating
hypersurfaces be represented with a \textbf{tilde} on them, as in $\tilde{e}$.
If we want to refer to the both types of these edges by a single variable,
then we use the $e$ notation as before. I assume the same conventions apply
for bones also.
\end{notation}

Consider the expression for the partition function:%
\begin{equation}
Z=\sum_{\left\{  J_{b}\right\}  }\left[  \left(  \prod_{b}d_{J_{b}}\right)
Tr\left(  \int\prod_{b}\bigotimes_{b\subset e}\rho_{J_{b}}(g_{e})\prod
_{e}dg_{e}\right)  \right]
\end{equation}
Let us do the integration in the $g_{\hat{e}}$ variables of the edges $\hat
{e}$ that lie on the foliating surfaces only. Then the product of the edge
integrals of these edges in the above equation is replaced by a product of the
intertwiners. The resulting integrand in the right hand side of the above
equation is made up of a product of spin network functionals \cite{RS} with
parallel propagators constructed out of certain products of the $\rho
_{J_{\hat{b}}}(g_{\tilde{e}})$'s and the intertwiners $i_{\hat{e}}$'s
intertwining them. In figure (\ref{fig.spinder}) this process has been
explained and many of the notations are illustrated in $1+1$ dimensions.%

There are two spin network functionals for each $\Omega_{i}$. One of them,
$\psi_{i}^{+}$(the other is $\psi_{i}^{-}$) is made up of the intertwiners
associated with the sides of all the edges $\hat{e}$ of $\Sigma_{i}$ facing
$\Omega_{i}$ ($\Omega_{i-1}$) and the $\rho_{J_{\hat{b}}}($ $g_{\tilde{e}})$'s
of the edges $\tilde{e}$ in $\Omega_{i}$($\Omega_{i-1}$). These spin network
functionals will be explained in more detail next.

\begin{figure}
[ptbh]
\begin{center}
\includegraphics[
height=4.5109in,
width=4.3621in
]%
{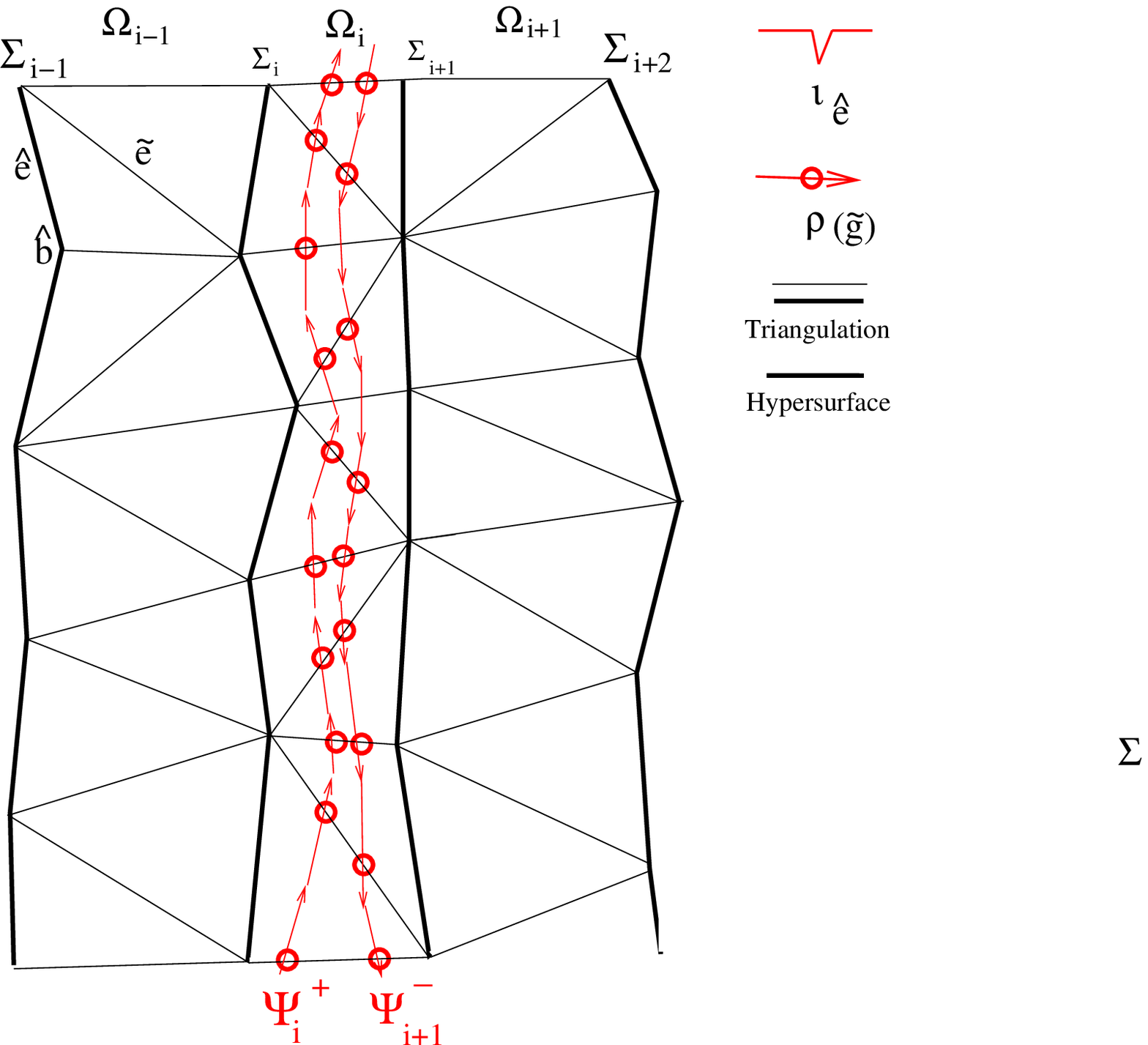}%
\caption{A foliation and the spin network functionals in 2D.}%
\label{fig.fol}%
\end{center}
\end{figure}

In figure (\ref{fig.fol}) the spin network functionals are shown.%

\section{The spin network functionals.}

To clearly see the various elements in $Z$, let us define a set of notations.

\begin{notation}
Let $\{g_{\tilde{e}}\}_{i}$ be the set of the $g_{\tilde{e}}$'s associated
with the edges $\tilde{e}$ in $\Omega_{i}$.
\end{notation}

\begin{notation}
Let $\bar{\Sigma}_{i}$ ($\bar{\Omega}_{i}$) be the triangulation dual to the
triangulation of $\Sigma_{i}$ ($\Omega_{i}$) for any $i$.
\end{notation}

The dual triangulation $\bar{\Sigma}_{i}$ serves as a graph to define spin
network functionals. For every edge and bone in the triangulation of $M$ in
$\Sigma_{i}$ there is a node and a link in the graph $\bar{\Sigma}_{i}$, respectively.

\begin{notation}
Let the coloring $\{J_{\hat{b}},o_{\hat{b}},i_{\hat{e}}\}_{\bar{\Sigma}}$ be
the set of representations $J_{\hat{b}}$'s, orientations $o_{\hat{b}}$'s
assigned to the bones $\hat{b}$'s and intertwiners $i_{\hat{e}}$ associated
with the edges $\hat{e}$'s of the manifold on any hypersurface $\Sigma$.
\end{notation}

Notice that the bones $\hat{b}$'s of the simplicial manifold $M$ are actually
the edges of the hypersurface on which they are lying. But we will refer to
the simplices as edges or bones with respect to the simplicial manifold $M$ to
keep our notations simple.

The orientation $o_{\hat{b}}$'s can be used to assign directions (arrows) to
the links dual to the bones $\hat{b}$'s. The arrows assigned to the links can
be used to restrict the choice of the intertwiner $i_{\hat{e}}$ assigned to
the node corresponding to the edge $\hat{e}$ $\subset\Sigma_{i}$, as a linear
map from the tensor product of the representations assigned to the links with
incoming arrows converging at the node, to the tensor product of the
representations assigned to the links with outgoing arrows diverging out of
the node. If all the arrows are incoming or outgoing then the intertwiner
linearly maps to or from the identity representation, respectively.

\begin{definition}
Given a hypersurface $\Sigma_{i}$ and any bone $\hat{b}$ on it, we can
associate parallel propagators $G_{\hat{b}}^{\pm}$ to the bone, defined as
follows:
\begin{equation}
G_{\hat{b}}^{+}=\prod\limits_{\forall\tilde{e}\backepsilon\hat{b},\tilde{e}%
\in\Omega_{i}}g_{\tilde{e}}. \label{gbone1}%
\end{equation}%
\begin{equation}
G_{\hat{b}}^{-}=\prod\limits_{\forall\tilde{e}\backepsilon\hat{b},\tilde{e}%
\in\Omega_{i-1}}g_{\tilde{e}}. \label{gbone2}%
\end{equation}
where, the multiplications are done in the sequential order of edges
$\tilde{e}\ni\hat{b}$ around the bone $\hat{b}$. The starting edge for
$G_{\hat{b}}^{+}$($G_{\hat{b}}^{-}$) is given by orientation $o_{\hat{b}}%
$($\bar{o}_{\hat{b}}$ =opposite orientation to $o_{\hat{b}}$). Let the
collection of these $G_{\hat{b}}^{\pm}$'s for all $\hat{b}$ on $\Sigma_{i}$ be
denoted as $\{G_{\hat{b}}^{\pm}\}_{\Sigma_{i}}$.
\end{definition}

\begin{figure}
[ptbh]
\begin{center}
\includegraphics[
height=2.9542in,
width=3.6469in
]%
{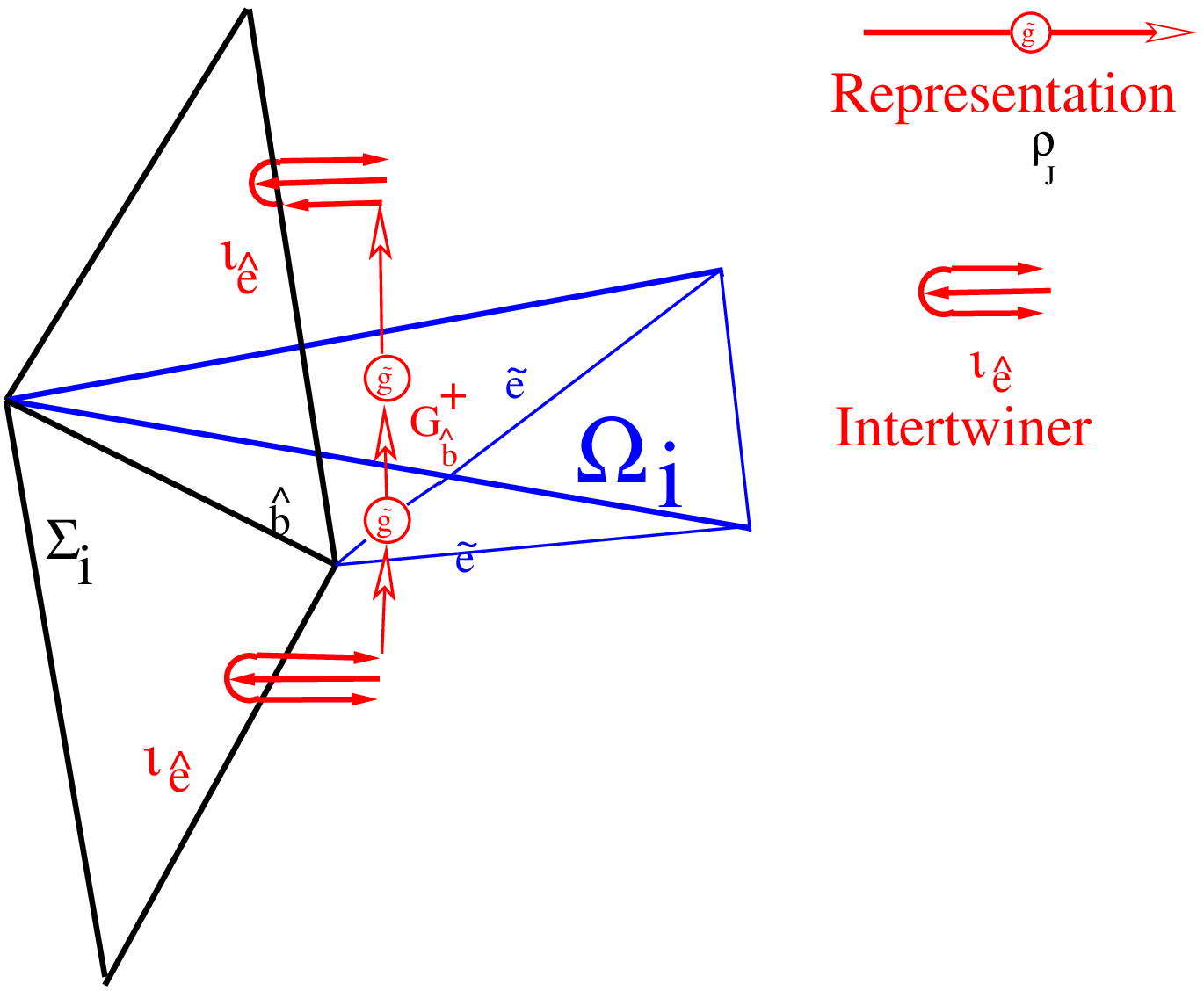}%
\caption{A $\rho_{J}(G_{\hat{b}}^{+})$ intertwined between two intertwiners.}%
\label{fig.gbone}%
\end{center}
\end{figure}

In figure (\ref{fig.gbone}) a $\rho_{J}(G_{\hat{b}}^{+})$ intertwined between
two intertwiners is shown.%

\begin{notation}
Let $X_{i_{\hat{e}}}$ be the extra free variable that uniquely fixes the
intertwiner $i_{\hat{e}}$, given the orientations $o_{\hat{b}}$'s and
$J_{\hat{b}}$'s associated with the bones $\hat{b}$ $\in\hat{e}$.
\end{notation}

\begin{definition}
By associating $\{G_{\hat{b}}^{+}\}_{\Sigma_{i}}$ to the links of $\bar
{\Sigma}_{i}$ and the intertwiners $\{i_{\hat{e}},\hat{e}\in\Sigma_{i}\}$ to
the nodes of $\bar{\Sigma}_{i}$, and tracing them according to the topology of
and orientations of links in $\bar{\Sigma}_{i}$, we can define spin network
functionals \cite{RS} associated with $\bar{\Sigma}_{i}$. I multiply this by a
\textbf{normalizing factor}\footnote{The bone amplitude $d_{J_{\hat{b}}}$ of
the bones on the hypersurface $\Sigma_{i}$ has been equally factored between
the two sides of $\Sigma_{i-1}$ and $\Sigma_{i}$.This is responsible for this
${\prod_{\hat{b}\in\Sigma_{i}}}d_{J_{\hat{b}}}^{\frac{1}{2}}$ term.}
${\prod_{\hat{b}\in\Sigma_{i}}}d_{J_{\hat{b}}}^{\frac{1}{2}}$ and we denote
this by
\begin{equation}
\psi_{i}^{+}=\psi(\bar{\Sigma}_{i},\{J_{\hat{b}},o_{\hat{b}},X_{i_{\hat{e}}%
},G_{\hat{b}}^{+}\}_{\Sigma_{i}}). \label{eq.spin+}%
\end{equation}
This spin network functional associated with the side of $\Sigma_{i}$ that
faces $\Omega_{i}$. Similarly, by using $\{G_{\hat{b}}^{-}\}_{\Sigma_{i}}$ we
can define another spin network functional%
\begin{equation}
\psi_{i}^{-}=\psi(\bar{\Sigma}_{i},\{J_{\hat{b}},\bar{o}_{\hat{b}}%
,X_{\bar{\imath}_{\hat{e}}},G_{\hat{b}}^{-}\}_{\Sigma_{i}}). \label{eq.spin-}%
\end{equation}
associated with the other side of $\Sigma_{i}$ that faces $\Omega_{i-1}$. In
$\psi_{i}^{-}$we have used the adjoints $\bar{\imath}_{\hat{e}}$'s of the
intertwiners $i_{\hat{e}}$'s and opposite orientations $\bar{o}_{\hat{b}}$'s
to $o_{\hat{b}}$'s. These spin network functionals so defined capture the
gauge invariant information in the discretized connection $\{g_{\tilde{e}%
},\tilde{e}\in\Omega_{i}\}$.
\end{definition}

\begin{figure}
[ptbh]
\begin{center}
\includegraphics[
height=3.4186in,
width=4.0119in
]%
{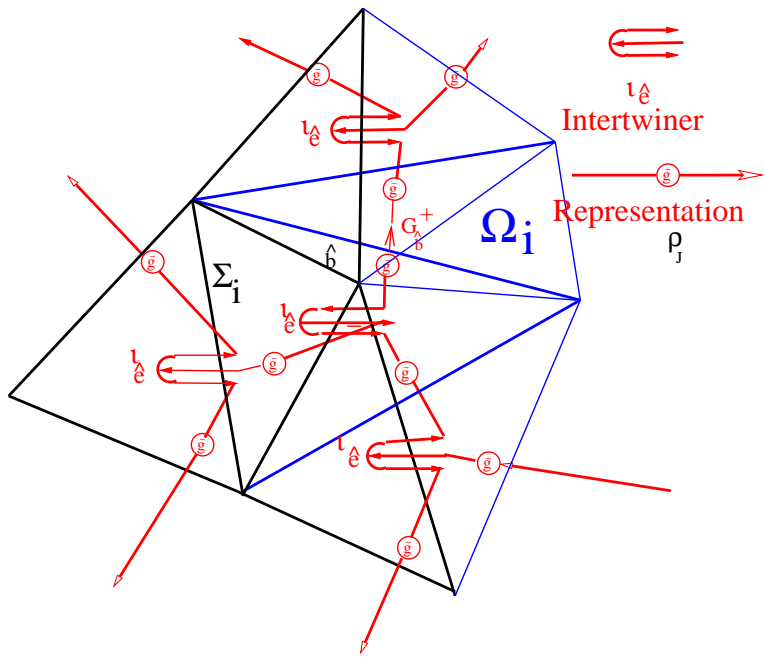}%
\caption{The spin network $\psi_{i}^{+}=\psi(\bar{\Sigma}_{i},\{J_{\hat{b}%
},o_{\hat{b}},i_{\hat{e}},G_{\hat{b}}^{+}\}_{\Sigma_{i}})$.}%
\label{fig.spin}%
\end{center}
\end{figure}

In figure (\ref{fig.spin}) part of $\psi_{i}^{+}=\psi(\bar{\Sigma}%
_{i},\{J_{\hat{b}},o_{\hat{b}},X_{i_{\hat{e}}},G_{\hat{b}}^{+}\}_{\Sigma_{i}%
})$ is shown graphically.%

\begin{definition}
\textit{ }Define an \textbf{inner product} between two spin network
functionals with two different colorings, associated with the same side of a
hypersurface.%
\begin{align*}
\left(  \psi,\psi^{\prime}\right)   &  \equiv\int\bar{\psi}\psi^{\prime}%
{\displaystyle\prod\limits_{\tilde{e}\in\Omega_{i}}}
dg_{\tilde{e}}\\
&  =\int\bar{\psi}(\bar{\Sigma},\{J_{\hat{b}},o_{\hat{b}},X_{i_{\hat{e}}%
},G_{\tilde{b}}\}_{\Sigma})\psi(\bar{\Sigma},\{J_{\hat{b}}^{^{\prime}}%
,o_{\hat{b}}^{\prime},X_{\acute{\imath}_{\hat{e}}},G_{\tilde{b}}\}_{\Sigma})%
{\displaystyle\prod\limits_{\tilde{e}\in\Omega}}
dg_{\tilde{e}}.
\end{align*}

\end{definition}

The $G_{\hat{b}}$'s are defined as same as in equation (\ref{gbone1}) or
equation (\ref{gbone2}) in relation to the bones of a boundary $\Sigma$ of
$\Omega$ and the parallel propagators $g_{\tilde{e}}$'s of $\Omega$.

\textit{It can be shown that these spin network functionals are
\textbf{orthonormal} in the inner product.}%
\begin{align*}
&  \left(  \psi,\psi^{\prime}\right) \\
&  =\int\bar{\psi}(\bar{\Sigma},\{J_{\hat{b}},o_{\hat{b}},X_{i_{\hat{e}}%
},G_{\tilde{b}}\}_{\Sigma})\psi(\bar{\Sigma},\{J_{\hat{b}}^{^{\prime}}%
,o_{\hat{b}}^{\prime},X_{\acute{\imath}_{\hat{e}}},G_{\tilde{b}}\}_{\Sigma})%
{\displaystyle\prod\limits_{\tilde{e}\in\Omega}}
dg_{\tilde{e}}\\
&  =%
{\displaystyle\prod_{\tilde{b}\in\Sigma}}
\delta_{J_{\hat{b}}J_{\hat{b}}^{\prime}}%
{\displaystyle\prod\limits_{\tilde{e}\in\Omega}}
\delta_{X_{i_{\hat{e}}}X_{\acute{\imath}_{\hat{e}}^{\prime}}}%
{\displaystyle\prod\limits_{\hat{b}\mid J_{\hat{b}}\ncong\bar{J}_{\hat{b}}}}
\delta_{o_{\hat{b}}o_{\hat{b}}^{\prime}}\text{,}%
\end{align*}
where the last product in above equation is only over bones whose associated
representations are not conjugate equivalent. This is because in case the
representation $J_{\hat{b}}$ is conjugate equivalent then the spin network
state does not change under the change of orientation $o_{\hat{b}}$, in other
words, for the corresponding state, $o_{\hat{b}}$ is physically redundant.

\begin{proposition}
The spin network functionals $\psi_{i}^{+}$ and $\psi_{i}^{-}$ are gauge invariant.
\end{proposition}

\begin{proof}
Now let us demonstrate the gauge invariance of the spin network functionals
those we obtained from the $BF$ spin foam evaluation. Let us gauge transform
the discrete connection $\{g_{e},e\in M\}$. This requires associating a gauge
transformation matrix $t_{s}$ $\in$ $G$ to each $n$-simplex $s$. Let us denote
the two simplices between which a edge $e$ of $M$ lies, as $s_{e,1}$ and
$s_{e,2}$, the numbers $1$ and $2$ are chosen such that the parallel
propagator $g_{e}$ propagates vectors from $s_{e,1}$ to $s_{e,2}$. After the
gauge transformation, the new discrete connection is $\{g_{e}^{\prime},e\in
M\}$, where $g_{e}^{\prime}=t_{s_{e,2}}g_{e}t_{s_{e,1}}^{-1}$. Let us denote
the two edges between which each bone $\hat{b}\in\Sigma_{i}$ lies as $\hat
{e}_{\hat{b},1}$ and $\hat{e}_{\hat{b},2}$, the numbers $1$ and $2$ are chosen
such that the orientation $o_{\hat{b}}$ points from $\hat{e}_{\hat{b},1}$ to
$\hat{e}_{\hat{b},2}$. Now if $s_{\hat{e}_{\hat{b},1}}$ and $s_{\hat{e}%
_{\hat{b},2}}$ are the simplices of $\Omega_{i}$ that touch $\Sigma_{i}$ at
$\hat{e}_{\hat{b},1}$ and $\hat{e}_{\hat{b},2}$ respectively, then under gauge
transformation we have $G_{\hat{b}}^{+}$ become $G_{\hat{b}}^{\prime
+}=t_{s_{\hat{e}_{\hat{b},1}}}G_{\hat{b}}^{+}t_{s_{\hat{e}_{\hat{b},2}}}^{-1}$
and our $\psi_{i}^{+}$ transforms to $\psi_{i}^{\prime+}=\psi(\bar{\Sigma}%
_{i},\{J_{\hat{b}},o_{\hat{b}},X_{i_{\hat{e}}},G_{\hat{b}}^{\prime+}%
\}_{\Sigma_{i}})$. But since we have traced the $G_{\hat{b}}^{\prime+}$ with
the intertwiners and the intertwiners are invariant under the action of group
elements of $G$, the gauge transformation matrices $t_{s}^{\prime}s$ are
absorbed by the intertwiners and we get back the original state. This proves
the gauge invariance of $\psi_{i}^{+}$. The gauge invariance of $\psi_{i}^{-}$
can be proved in the similar way.
\end{proof}

\begin{definition}
Let us define a functional $\left\langle \psi_{i}^{+},\psi_{i+1}%
^{-}\right\rangle $ as follows,
\begin{equation}
\left\langle \psi_{i}^{+},\psi_{i+1}^{-}\right\rangle _{\bar{\Omega}_{i}}%
=\int\psi_{i}^{+}\psi_{i+1}^{-}{\prod_{\tilde{b}\in\Omega_{i}}}\delta
(H_{\tilde{b}}){\prod_{\tilde{e}\in\Omega_{i}}}dg_{\tilde{e}}. \label{eq.sc}%
\end{equation}
Now it is straight forward to show that $Z$ can be rewritten as
\begin{equation}
Z={\sum_{\{J_{\hat{b}},o_{\hat{b}},X_{i_{\hat{e}}}\}}}{\prod_{i}}\left\langle
\psi^{+},\psi_{i+1}^{-}\right\rangle _{\bar{\Omega}_{i}},
\end{equation}
where $\{J_{\hat{b}},o_{\hat{b}},X_{i_{\hat{e}}}\}$ is the collection of
$\{J_{\hat{b}},o_{\hat{b}},X_{i_{\hat{e}}}\}_{i}$ for all $i$.
\end{definition}

\section{The elementary transition amplitudes}

Let us first fix a triangulation $\bar{M}$ of the manifold $M$ that satisfies
the properties enlisted in the previous section. Let us first calculate a
connection to connection transition amplitude using the path integral
formulation of quantum mechanics. The order of the hypersurfaces $i$ can be
considered to define a discrete coordinate time variable.

\begin{notation}
Let $\{g_{\tilde{e}}\}_{A}$ ($\{g_{\tilde{e}}\}_{B}$) associated with
$\Omega_{A}$ ($\Omega_{B}$) be the initial (final) connection information,
where $A$ and $B$ are integers such that $A<B$.
\end{notation}

\begin{notation}
Let $\Omega_{AB}$ be the simplicial manifold between $\Sigma_{A}$ and
$\Sigma_{B}$.
\end{notation}

\begin{definition}
A transition amplitude from $\{g_{\tilde{e}}\}_{A}$ to $\{g_{\tilde{e}}\}_{B}$
can be defined based on the path integral formulation as follows,%
\begin{equation}
\left\langle \{g_{\tilde{e}}\}_{A}|\{g_{\tilde{e}}\}_{B}\right\rangle
=\int\exp[iS_{AB}]{\prod_{e\in\Omega_{A+1,B-1}}}dg_{e}{\prod_{b\in\Omega_{AB}%
}}dB_{b}, \label{eq.path}%
\end{equation}
where $S_{AB}$ is defined to be%
\[
S_{AB}=\sum_{b\in\Omega_{A+1,B-1}}Tr(B_{b}\ln H_{b}).
\]
Our definition of the transition amplitudes has been chosen such that it satisfies:
\end{definition}

\begin{itemize}
\item the relationship%
\begin{align*}
\left\langle \{g_{\tilde{e}}\}_{A}|\{g_{\tilde{e}}\}_{C}\right\rangle  &
=\int\left\langle \{g_{\tilde{e}}\}_{A}|\{g_{\tilde{e}}\}_{B}\right\rangle \\
&  \left\langle \{g_{\tilde{e}}\}_{B}|\{g_{\tilde{e}}\}_{C}\right\rangle
{\prod_{\tilde{b}\in\Omega_{B}}}\delta(H_{b}){\prod_{\tilde{e}\in\Omega_{B}}%
}dg_{\tilde{e}},
\end{align*}
where the $A$, $B$ and $C$ are three consecutive integers in the increasing order,

\item and leads to the result,%
\begin{equation}
\left\langle \{g_{\tilde{e}}\}_{A}|\{g_{\tilde{e}}\}_{B}\right\rangle
={\sum_{\{J_{\hat{b}},o_{\hat{b}},X_{i_{\hat{e}}}\}}\psi_{A+1}^{-}%
\prod_{i=A+1}^{B-1}}\left(  \psi_{i}^{+},\psi_{i+1}^{-}\right)  \psi_{B}^{+}
\label{eq.path1}%
\end{equation}
on integration over $g_{\tilde{e}}$'s.
\end{itemize}

The above two properties can be checked explicitly by calculations.

For BF theory the physical states require only flat connections. If the
$g_{\tilde{e}}$ are restricted to flat connections $g_{\tilde{e}}^{f}$
($H_{\tilde{b}}=1$), then $\delta(H_{b})$ can be removed in the definition of
the transition amplitudes\footnote{An appropriate measure factor may need to
be introduced in the integrand.}. Then the first condition simply expresses an
abstract quantum mechanical property that any quantum transition amplitudes
have to satisfy:%
\begin{align*}
\left\langle \{g_{\tilde{e}}^{f}\}_{A}|\{g_{\tilde{e}}^{f}\}_{C}\right\rangle
&  =\int\left\langle \{g_{\tilde{e}}^{f}\}_{A}|\{g_{\tilde{e}}^{f}%
\}_{B}\right\rangle \\
&  \left\langle \{g_{\tilde{e}}^{f}\}_{B}|\{g_{\tilde{e}}^{f}\}_{C}%
\right\rangle {\prod_{\tilde{e}\in\Omega_{B}}}dg_{\tilde{e}}^{f}.
\end{align*}
This tells us that the transition amplitudes are intuitively well
defined\textit{.}

\begin{definition}
Let us define $S=\left\{  |\{J_{\hat{b}},o_{\hat{b}},X_{i_{\hat{e}}}%
\}_{\bar{\Sigma}},\bar{\Sigma}>\right\}  $ to be an orthonormal basis of
quantum states, each quantum state identified by a graph $\bar{\Sigma}$ and
the associated coloring of bones and edges by $\{J_{\hat{b}},o_{\hat{b}%
},X_{i_{\hat{e}}}\}_{\bar{\Sigma}}$. The $S$ defines a basis of abstract spin
network states \cite{RS}.
\end{definition}

\begin{definition}
Equation (\ref{eq.path1}) suggests the following two interpretations:
\end{definition}

\begin{enumerate}
\item The functional ${\psi_{i+1}^{-}}$ is the connection $\{g_{\tilde{e}%
}\}_{i}$ to the spin network state $\left\vert J_{\hat{b}},o_{\hat{b}%
},X_{i_{\hat{e}}}\}_{\bar{\Sigma}_{i+1}},\bar{\Sigma}_{i+1}\right\rangle $
transition amplitude,%
\[
{\psi_{i+1}^{-}=}\psi(\bar{\Sigma}_{i+1},\{J_{\hat{b}},\bar{o}_{\hat{b}%
},X_{\bar{\imath}_{\hat{e}}},G_{\hat{b}}^{-}\}_{\Sigma_{i+1}})=\left\langle
\{g_{\tilde{e}}\}_{i}|\{J_{\hat{b}},o_{\hat{b}},X_{i_{\hat{e}}}\}_{\bar
{\Sigma}_{i+1}},\bar{\Sigma}_{i}\right\rangle _{\bar{\Omega}_{i}},
\]
and similarly, functional ${\psi_{i}^{+}}$ is the spin network state
$|\{J_{\hat{b}},o_{\hat{b}},X_{i_{\hat{e}}}\}_{i},\bar{\Sigma}_{i}>$ to the
connection $\{$ $g_{\tilde{e}}\}_{i}$ transition amplitude%
\[
{\psi_{i}^{+}=}\psi(\bar{\Sigma}_{i},\{J_{\hat{b}},o_{\hat{b}},X_{i_{\hat{e}}%
},G_{\bar{b}}^{+}\}_{\Sigma_{i}})=\left\langle \{J_{\hat{b}},o_{\hat{b}%
},X_{i_{\hat{e}}}\}_{\bar{\Sigma}_{i}},\bar{\Sigma}_{i}|\{g_{\tilde{e}}%
\}_{i}\right\rangle _{\bar{\Omega}_{i}}%
\]
Please notice that the above equations are consistent with the
orthonormalities of $\psi_{i}^{\pm}$ and the $|\{J_{\hat{b}},o_{\hat{b}%
},X_{i_{\hat{e}}}\}_{\bar{\Sigma}},\bar{\Sigma}>$'s. I have the suffix
$\bar{\Omega}_{i}$ in the left-hand sides of the last two equations indicate
the dependence on $\bar{\Omega}_{i}$.
\end{enumerate}

\begin{enumerate}
\item[2.] Our $\left\langle \psi_{i}^{+},\psi_{i+1}^{-}\right\rangle $ is the
spin network state to spin network state transition amplitude \footnote{Using
\ref{eq.sc} we can show that the elementary transition amplitude $\left\langle
\psi_{i}^{-},\psi_{i}^{+}\right\rangle $ is simply the product of the quantum
amplitudes of the $n$-simplices in $\Omega_{i}$, of the bones $\hat{b}$
$\in\Omega_{i}$ and the square root of the quantum amplitudes of the bones on
$\Sigma_{i}$ and $\Sigma_{i+1}$.},%
\begin{equation}
\left\langle \psi_{i}^{+},\psi_{i+1}^{-}\right\rangle =\left\langle
\{J_{\hat{b}},o_{\hat{b}},X_{i_{\hat{e}}}\}_{\bar{\Sigma}_{i}},\bar{\Sigma
}_{i}|\{J_{\hat{b}},o_{\hat{b}},X_{i_{\hat{e}}}\}_{\bar{\Sigma}_{i+1}}%
,\bar{\Sigma}_{i+1}\right\rangle _{\bar{\Omega}_{i}} \label{eq.trans}%
\end{equation}
Equation(\ref{eq.trans}) defines an elementary transition amplitude. The
suffix has been added in the right side above equation because the elementary
transition amplitude depends on the triangulation of $\bar{\Omega}_{i}$.
\end{enumerate}

\begin{definition}
The graphs $\bar{\Sigma}_{i}$ and $\bar{\Sigma}_{i+1}$ do not uniquely
determine $\bar{\Omega}_{i}$. To remove the triangulation dependence of
$\left\langle \psi_{i}^{+},\psi_{i+1}^{-}\right\rangle _{\Omega_{i}}$, let us
define a new elementary transition amplitude, by summing over all possible
one-simplex thick triangulations of $\bar{\Omega}_{i}$ that sandwich between
$\bar{\Sigma}_{i}$ and $\bar{\Sigma}_{i+1}$ as follows:%
\begin{align}
& \left\langle \{J_{\hat{b}},o_{\hat{b}},X_{i_{\hat{e}}}\}_{\bar{\Sigma}_{i}%
},\bar{\Sigma}_{i}|\{J_{\hat{b}},o_{\hat{b}},i_{\hat{e}}\}_{i+1},\bar{\Sigma
}_{i+1}\right\rangle \label{eq.sc1}\\
& =\sum_{\bar{\Omega}_{i}}\left\langle \{J_{\hat{b}},o_{\hat{b}},X_{i_{\hat
{e}}}\}_{\bar{\Sigma}_{i}},\bar{\Sigma}_{i}|\{J_{\hat{b}},o_{\hat{b}%
},X_{i_{\hat{e}}}\}_{\bar{\Sigma}_{i+1}},\bar{\Sigma}_{i+1}\right\rangle
_{\bar{\Omega}_{i}}.\nonumber
\end{align}

\end{definition}

\begin{proposition}
Given any two abstract spin network states $\left\vert \{J_{\hat{b}}%
,o_{\hat{b}},X_{i_{\hat{e}}}\}_{\bar{\Sigma}_{A}},\bar{\Sigma}_{A}%
\right\rangle $ and $\left\vert \{J_{\hat{b}},o_{\hat{b}},X_{i_{\hat{e}}%
}\}_{\bar{\Sigma}_{B}},\bar{\Sigma}_{B}\right\rangle $, one can immediately
calculate an elementary transition amplitude between them.
\end{proposition}

\begin{proof}
The new elementary transition amplitude defined in equation (\ref{eq.sc1})
depends only on the $(\{J_{\hat{b}},o_{\hat{b}},X_{i_{\hat{e}}}\}_{\bar
{\Sigma}_{i}}$,$\bar{\Sigma}_{i})$ and the $(\{J_{\hat{b}},o_{\hat{b}%
},X_{i_{\hat{e}}}\}_{\bar{\Sigma}_{i+1}},\bar{\Sigma}_{i+1})$. Because of
this, given any two spin network states $\left\vert \{J_{\hat{b}},o_{\hat{b}%
},X_{i_{\hat{e}}}\}_{\bar{\Sigma}_{A}},\bar{\Sigma}_{A}\right\rangle $ and
$\left\vert \{J_{\hat{b}},o_{\hat{b}},X_{i_{\hat{e}}}\}_{\bar{\Sigma}_{B}%
},\bar{\Sigma}_{B}\right\rangle $, one can immediately calculate a transition
amplitude between them, by constructing a one simplex-thick simplicial
manifold $\Omega_{AB}$ that sandwiches between $\Sigma_{A}$ and $\Sigma_{B}$,
constructing the functions $\psi_{A}^{-}(\Sigma_{A},\{J_{\hat{b}},\bar
{o}_{\hat{b}},X_{\bar{\imath}_{\hat{e}}},G_{\tilde{e}}^{-}\}_{\Sigma_{A}})$
and $\psi_{B}^{+}(\Sigma_{B},\{J_{\hat{b}},o_{\hat{b}},X_{i_{\hat{e}}%
},G_{\tilde{e}}^{+}\}_{\Sigma_{B}})$, using equation (\ref{eq.sc1}) to
calculate $\left\langle \{J_{\hat{b}},o_{\hat{b}},X_{i_{\hat{e}}}%
\}_{\bar{\Sigma}_{A}},\bar{\Sigma}_{A}|\{\bar{J}_{\hat{b}},X_{\bar{\imath
}_{\hat{e}}}\}_{\bar{\Sigma}_{B}},\bar{\Sigma}_{B}\right\rangle $. If there is
no $\Omega_{AB}$ that fit between $\Sigma_{A}$ and $\Sigma_{B}$ then the
elementary transition amplitude has to be defined to be zero.
\end{proof}

The elementary transition amplitudes defined above can be further generalized
as follows. Let $H$ be an abstract Hilbert space linearly spanned by the spin
network state basis defined earlier, $S$ then for any two $\left\vert
\psi\right\rangle ,\left\vert \phi\right\rangle $ $\in$ $H$, the transition
amplitude $\left\langle \psi|\phi\right\rangle $ can be defined by extending
the elementary transition amplitude by linearity. So our elementary transition
amplitudes defines a transition matrix. Then if the index $i$ is considered to
represent a coordinate time, the transition matrix evolves any state
$\left\vert \psi_{i}\right\rangle \in H$ at a discrete time instant $i$ to its
next time instant $i+1$. In the case of an arbitrary group $G$ BF theory $i$
may be just an arbitrary parameter to help explore its quantum theory.
\emph{But in the case of the Lorentzian quantum gravity, the index }$i$\emph{
does have some physical relation to time}. (Please see the discussion near the
end of the section on the $3+1$ Formulation of Gravity). \emph{The elementary
transition matrix }$\left\langle \{J_{\hat{b}},o_{\hat{b}},X_{i_{\hat{e}}%
}\}_{\bar{\Sigma}_{i}},\bar{\Sigma}_{i}|\{J_{\hat{b}},o_{\hat{b}}%
,X_{i_{\hat{e}}}\}_{\bar{\Sigma}_{i+1}},\bar{\Sigma}_{i+1}\right\rangle
$\emph{ so defined helps define a discrete co-ordinate time evolution scheme
of BF theory. }In section six, we will explain how to adapt this scheme to
gravity by redefining the elementary transition amplitudes.

A close analysis indicates that the \textbf{topology change} is built into
this formalism. Please see the section on $1+1$ BF theory for an illustration.

Our spin network functionals in four dimensions for the BF theory and those
for gravity that will be discussed later are similar to those in canonical
quantum gravity on a triangulated three manifold formulated by Thiemann
\cite{TT1}, \cite{R3}. In Thiemann's formulation, the spin networks are
constructed using parallel propagators associated with the edges of the
three-simplices of a triangulation of a three manifold. These parallel
propagators are constructed out of the path ordered integral $P\exp(-\int A)$
of the Ashtekar-Sen connection \cite{As} on the manifold. Our spin network
functionals are constructed using the parallel propagators $g_{\tilde{e}}$
associated with the edges $\tilde{e}$ of the four-simplices in the four
dimensional slices $\Omega_{i}$. The four dimensional slices $\Omega_{i}$ can
be considered as thickened $3D$ simplicial surfaces. In our formulation the
physical meaning of the parallel propagators $g$'s is clear.

Further work that needs to be done on the theoretical constructions developed
in this section will be discussed at the end of this article.

\section{The $1+1$ splitting of the 2D gravity.}

In $1+1$ dimensions the spin network functionals are mathematically simple.
Here the 2D manifold is foliated by 1D curves. To simplify our discussion, let
us restrict ourselves to conjugate equivalent representations, but conjugate
inequivalent representations can be easily included by adding additional delta
functions in the transition amplitude calculations.

\subsection{The one circle to one circle elementary transition amplitude.}

Assume $\Omega_{i}$ for a given $i$ is topologically a cylinder. This means
that $\Sigma_{i}$ and $\Sigma_{i+1}$ are topologically circles.%

\begin{figure}
[ptbh]
\begin{center}
\includegraphics[
trim=0.011250in 0.000000in -0.011251in 0.000000in,
height=1.0931in,
width=4.5325in
]%
{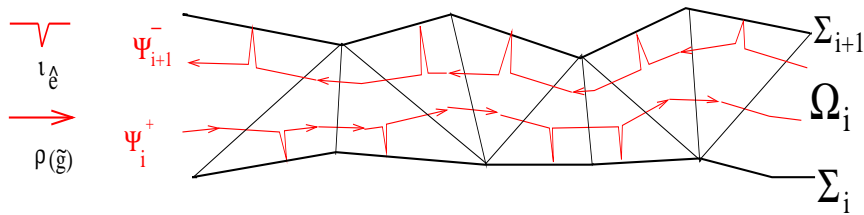}%
\caption{The 2D foliation.}%
\label{fig.2dfol}%
\end{center}
\end{figure}
In figure(\ref{fig.2dfol}), two consecutive foliating hypersurfaces,
$\Omega_{i}$ and the spin networks functionals in between them are shown. Only
part of the cylinder has been shown. The intertwiners are given by
equation$\left(  \ref{eq.2int}\right)  $. The $\delta_{J_{\hat{b}_{1}}%
J_{\hat{b}_{2}}}$ term in the intertwiners specifies that the $J_{\hat{b}}$
are the same for all $\hat{b}$ belonging to a hypersurface $\Sigma_{i}$. Let
it be $J_{i}$ ($J_{i+1}$) for $\Sigma_{i}$ ( $\Sigma_{i+1}$). Let us assume
the $o_{\hat{b}}$ are same for all the bones on each circle of the foliations
and we will comment on more general cases later. Let each $\Sigma_{i}$ (
$\Sigma_{i+1}$) be made of $N_{i}$ ( $N_{i+1}$) edges. Using expressions for
the intertwiners given in equation (\ref{eq.2int}), the spin network
functionals can be calculated as
\[
\psi_{i}^{+}=Tr(\rho_{J_{i}}(\prod_{\left\{  \tilde{e}\right\}  \in\Omega_{i}%
}g_{\tilde{e}}))
\]
and
\[
\psi_{i+1}^{-}=Tr(\rho_{J_{i+1}}(\prod_{\tilde{e}\in\Omega_{i}}g_{\tilde{e}%
})),
\]
where the powers of $d_{J}^{\frac{1}{2}}$ in $\psi_{i}^{+}$, $\psi_{i+1}^{-}$
in equation (\ref{eq.spin+}) and equation (\ref{eq.spin-}) from the bone
amplitudes and intertwiners cancel each other. The $\left\{  g_{\tilde{e}%
}\right\}  _{i}$ are multiplied according to the order defined by the
topological continuity of $\Omega_{i}$ and orientation $\bar{o}_{i}$. But
since we restricted ourselves to conjugate equivalent representations, the
spin network functionals are independent of the orientations $o_{\bar{b}}$.

In the $1+1$ formalism there is no internal holonomy between the foliations.
The elementary transition amplitudes can be calculated using equation $\left(
\ref{eq.sc}\right)  $ as follows:%
\[
\left\langle \psi_{i}^{+},\psi_{i+1}^{-}\right\rangle =\int\psi_{i}^{+}%
\psi_{i+1}^{-}{\prod_{\left\{  \tilde{e}\right\}  \in\Omega_{i}}}dg_{\tilde
{e}}%
\]%
\begin{align*}
&  =\int Tr(\rho_{J_{i}}({\prod_{\left\{  \tilde{e}\right\}  \in\Omega_{i}}%
}g_{\tilde{e}}))Tr(\bar{\rho}_{J_{i+1}}({\prod_{\left\{  \tilde{e}\right\}
\in\Omega_{i}}}g_{\tilde{e}})){\prod_{\left\{  \tilde{e}\right\}  \in
\Omega_{i}}}dg_{\tilde{e}}\\
&  =\delta_{J_{i}J_{i+1}},
\end{align*}
where $M_{i}$ is the number of edges in $\Omega_{i}$. It is interesting to see
that the elementary transition amplitude does not depend on the triangulation.

\subsection{The $n$-circle to $m$-circle elementary transition amplitude}

\begin{figure}
[ptbh]
\begin{center}
\includegraphics[
height=3.0009in,
width=2.757in
]%
{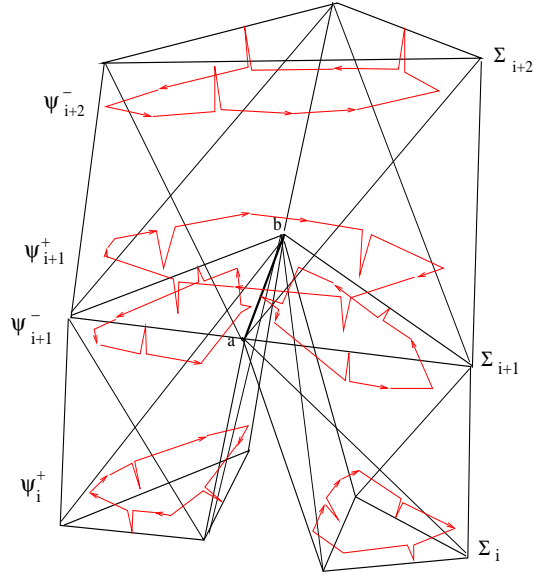}%
\caption{A topology change.}%
\label{1to2fig}%
\end{center}
\end{figure}

The case where a two-manifold transforms from two circle topology to
one-circle topology is shown in figure (\ref{1to2fig}). The triangulation
makes the circles look like triangles.

The spin network functionals $\psi_{i+2}^{-}$ and $\psi_{i+1}^{+}$ are exactly
same as in the previous section. Therefore the transition amplitude between
them is the same as before, $\delta_{J_{i+1}J_{i+2}}$. The $\psi_{i}^{+}$ is
made of a product of two one-circle spin network functionals.

The $\psi_{i+1}^{-}$ is also a product of two one-circle spin network
functionals except that it is missing a factor of $d_{j+1}$. This is because
the $d_{j+1}^{\frac{1}{2}}$ factors from the quantum amplitudes of the bones
and the intertwiners do not completely cancel each other. So when the
transition amplitude between $\psi_{i+1}^{-}$ and $\psi_{i}^{+}$ is calculated
we get a result of $d_{J_{i+1}}^{-1}$:%
\begin{align*}
\left\langle \psi_{i}^{+},\psi_{i+1}^{-}\right\rangle  &  =d_{j+1}%
^{-1}\left\langle \psi_{i}^{+},d_{j+1}\psi_{i+1}^{-}\right\rangle \\
&  =d_{j+1}^{-1}\delta_{J_{i}J_{i+1}}.
\end{align*}
The topologies of $\Sigma_{i+1}$ and $\Sigma_{i}$ are not the same. This
suggests that the above result is a quantum amplitude for a topology change
from one circle to two circles. The intertwiner of the edge \textit{ab} in
figure(\ref{1to2fig}) at which the two circles intersect contributed a factor
of $d_{j+1}^{-1}$ to the elementary transition amplitude.

\begin{figure}
[ptbh]
\begin{center}
\includegraphics[
height=2.898in,
width=2.5996in
]%
{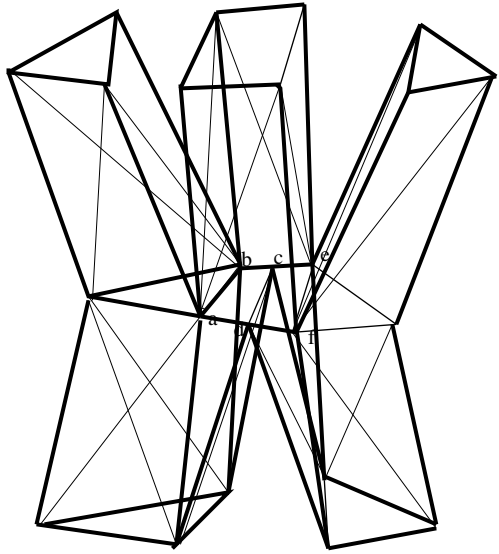}%
\caption{A more complicated topology change.}%
\label{2to3.fig}%
\end{center}
\end{figure}

In figure(\ref{2to3.fig}) a triangulation of a two-circle to three-circle
transforming 2-manifold is shown.%

There are three edges in this case at which the circles intersect each other.
Therefore the quantum transition amplitude here is $\delta_{J_{i}J_{i+1}%
}d_{J_{i}}^{-3}$.

In general a $n$-circle to $m$-circle changing two-manifold involves $n+m-2$
of these edges and so the elementary transition amplitude is $\delta
_{J_{i}J_{i+1}}d_{J_{i}}^{-\left(  n+m-2\right)  }$.

\begin{summary}
\textit{If the states of the two dimensional BF\ theory are represented by
}$\left\vert n,J\right\rangle $\textit{ where }$n$\textit{ is the number of
circles in the topology and }$J\ $\textit{is the representation for the spin
network states, then}%
\[
\left\langle m,J|n,K\right\rangle =\delta_{J_{i}J_{i+1}}d_{J_{i}}^{-(n+m-2)}.
\]

\end{summary}

It can be clearly seen from the above result that the transition matrix is
symmetric and is non-unitary.

\subsection{Topological invariance of the 2D transition amplitudes}

In case of 2D manifolds, the partition function is \cite{bz1}, \cite{AP}%
\begin{equation}
Z=\sum_{J}d_{J}^{\chi(M)} \label{eq2dpart}%
\end{equation}
where, $\chi$ is the Euler characteristic of $M$, a topological invariant.

Let $\Sigma_{A}$ and $\Sigma_{B}$ are two closed $1D$ manifolds. Let $M$ be a
simplicial $2$-manifold foliated by $N$ hypersurfaces $\{\Sigma_{i}\}$ such
that $\Sigma_{A}=\Sigma_{1}$ and $\Sigma_{B}=\Sigma_{N}$. Then the transition
amplitude $\left\langle \Sigma_{A},J_{A}|\Sigma_{B},J_{B}\right\rangle _{M}$
can be calculated by multiplying the elementary transition amplitudes
$\left\langle \Sigma_{i},J_{i}|\Sigma_{i+1},J_{i+1}\right\rangle $. Since for
$2$-manifolds the intertwiners require that all the $J_{b}$'s are the same we
have%
\[
\left\langle \Sigma_{A},J_{A}|\Sigma_{B},J_{B}\right\rangle _{M}=\delta
_{J_{A}J_{B}}%
{\displaystyle\prod\limits_{i=1}^{N-1}}
\left\langle \Sigma_{i},J_{A}|\Sigma_{i+1},J_{A}\right\rangle _{\Omega_{i}}.
\]

Now consider we have two copies of $M,$ and splice them at their identical
ends. Let the resultant manifold be $M^{^{\prime}}$. Then, we can show using
the calculations leading to equation(\ref{eq2dpart}) as done in \cite{bz1} and
\cite{AP} that $\left\langle \Sigma_{A},J_{A}|\Sigma_{B},J_{B}\right\rangle
_{M}\left\langle \Sigma_{B},J_{B}|\Sigma_{A},J_{A}\right\rangle _{M}$ is
nothing but the partition function associated with $M^{\prime}$ with $J$ fixed
to value $J_{A}$%
\begin{align*}
\left\langle \Sigma_{A},J_{A}|\Sigma_{B},J_{B}\right\rangle _{M}^{2}  &
=\left\langle \Sigma_{A},J_{A}|\Sigma_{B},J_{B}\right\rangle _{M}\left\langle
\Sigma_{B},J_{B}|\Sigma_{A},J_{A}\right\rangle _{M}\\
&  =d_{J_{A}}^{\chi(M^{\prime})}\delta_{J_{A}J_{B}}.
\end{align*}

Now, since the above result is a topological invariant, and the 2-D transition
amplitudes are always positive real, we can conclude $\left\langle \Sigma
_{A},J_{A}|\Sigma_{B},J_{B}\right\rangle _{M}$ is a topological invariant.
This means that the transition amplitude%
\[
\left\langle \Sigma_{A},J_{A}|\Sigma_{B},J_{B}\right\rangle =\sum
_{M}\left\langle \Sigma_{A},J_{A}|\Sigma_{B},J_{B}\right\rangle _{M}%
\]
where the summation is over all possible $M$ (an arbitrary triangulation for
each topology used) that sandwich between $\Sigma_{A}$ and $\Sigma_{B}$, is
independent of the triangulations of $M$.

The generalization of this result to higher dimensions is being analyzed and
will be published elsewhere.

\section{The $3+1$ formulation of gravity.}

Lets go the four dimensional cases after a brief discussion of the 3D
Riemannian case. The $3D$ Riemannian gravity is equivalent to the $3D$ BF
theory for the group $SU(2)$. The intertwiners are just the $3J$ symbols of
$SU(2)$. The spin network functionals are essentially the same as that of the
Ashtekar-Barbero Euclidean canonical quantum gravity formalism \cite{Barb}.
Here the spin network functionals live on the two dimensional foliating surfaces.

In the case of the $SO(4)$ Riemannian gravity, the most popular proposal is
the Barrett-Crane model \cite{bc1}, which was derived by imposing the
Barrett-Crane constraints on the spin foam model of the $SO(4)$ BF theory. The
Barrett-Crane constraints are basically the discretized Plebanski constraints.

The Barrett-Crane constraints are implemented on the $SO(4)$ BF theory given
by equation $\left(  \ref{eq.6}\right)  $ by using the following
conditions\footnote{The model so obtained may differ from the Barrett-Crane
model by the amplitudes of the lower dimensional ($<4$) simplices. We believe
that the imposition of the Barrett-Crane constraints are not yet derived in a
way that can be rigorously related to any discretized form of the gravity
Lagrangian. Because of this the amplitudes of the lower dimensional simplices
are not yet fixed. For simplicity, here we assume that these quantum amplitude
are same as that of the BF\ spin foam model.}:

\begin{enumerate}
\item The $J_{b}$ are restricted to the simple representations of $SO(4)$
\cite{bc1}, \cite{rz1}.

\item The intertwiners are restricted to the Barrett-Crane intertwiners given
in equation $\left(  \ref{eq.bc1}\right)  $ \cite{bc1}.

Please see appendix C for the definitions of the simple representations and
the Barrett-Crane intertwiner.
\end{enumerate}

To simplify the calculation of the edge integrals, the directions of the
holonomies in the derivation of the spin foam model can be chosen as
illustrated in figure (\ref{fig.holdir}). The parallel sets of arrows indicate
the direction in which the holonomies are traversed through the edges of a
four-simplex. Please refer to appendix A and B\ for more information.

The spin network functionals $\psi_{i}^{\pm}=\psi^{\pm}(\Sigma_{i}%
,\{J_{\hat{b}},o_{\hat{b}},X_{i_{\hat{e}}},G_{\bar{b}}^{\pm}\}_{\Sigma_{i}})$
of the $SO(4)$ BF theory can be adapted to gravity by restricting the $J_{b}%
$'s to the simple representations and the intertwiners $i_{e}$ to the
Barrett-Crane intertwiners \cite{bc1} Let $h:S^{3}\rightarrow SU(2)$ be a
mapping and $\rho_{J}$ be the $J$ representation of $SU(2)$. Then the
Barrett-Crane intertwiner can be rewritten as (derived in appendix C)%
\[
i_{l_{1}r_{1}l_{2}r_{2}l_{3}r_{3}l_{4}r_{4}}^{J_{1}J_{2}\bar{J}_{3}\bar{J}%
_{4}}=\int_{S^{3}}dx\rho_{r_{1}J_{1}}^{l_{1}}(h(x))\rho_{r_{2}J_{2}}^{l_{2}%
}(h(x))\rho_{l_{3}J_{3}}^{r_{3}}(h^{-1}(x))\rho_{l_{4}J_{4}}^{r_{4}}%
(h^{-1}(x)).
\]

\begin{definition}
The elementary transition amplitudes $\left(  \psi_{i}^{+},\psi_{i+1}%
^{-}\right)  $ defined in equation (\ref{eq.sc}) can be reformulated for
Riemannian gravity as
\begin{equation}
\left(  \psi_{i}^{+},\psi_{i+1}^{-}\right)  =\tilde{P}_{BC}\int\psi_{i}%
^{+}\psi_{i+1}^{-}{\prod_{\tilde{b}\in\Omega_{i}}}\delta(H_{\tilde{b}}%
){\prod_{\tilde{e}\in\Omega_{i}}}dg_{\tilde{e}}, \label{eq.gr.tr}%
\end{equation}
where $\tilde{P}_{BC}$ is the projector which imposes the Barrett-Crane
constraints on the intertwiners associated with the edges $\tilde{e}$.
\end{definition}

Any three-simplicial hypersurface $\Sigma$ with the $J$'s interpreted as the
sizes of the edges of its three-simplices, which are assumed to be flat
\cite{Rg2}, describes a discrete geometry. In this sense the above equation
assigns quantum amplitudes for a history of geometries \cite{bz1}.

In the case of Riemannian gravity the final spin network functional has been
constructed on the homogenous space $S^{3}=SO(4)/SU(2)$ corresponding to the
subgroup $SU(2)$.

In the case of $SO(3,1)\approx$ $SL(2,C)$, imposing the Barrett-Crane
constraints can potentially lead to three different types of spin foam models
relating to the three different homogenous space of $SO(3,1)$ corresponding to
the subgroups $SO(3)$, $SU(1,1)$ or $E(2)$ \cite{bc2}. The first case has been
more investigated than the other two and is the most interesting in the
context of our $3+1$ formulation. In this case, the theory is defined
\cite{bc2} by replacing $S^{3}$ in the Euclidean formalism defined above by
$H^{+}$ the homogenous space $SL(2,C)/SU(2)$. $H^{+}$ is the space of the
upper sheet of the two-sheeted hyperboloid of $4D$ Minkowski space-time. The
related spin network functional of the $3+1$ formulation is made of the
infinite dimensional representations of the Lorentz group. Here the $J_{b}$
values are continuous (more precisely, imaginary). An element $x$ of $H^{+}$,
is assigned to each side of each edge of the 4-simplices. The asymptotic limit
\cite{BS} of the theory is controlled by the Einstein-Regge action \cite{Rg2}
of gravity \cite{BS}. In the asymptotic limit the dominant contribution
(non-degenerate sector) to the spin foam amplitude is when the $x$ values are
normals to the edges in the simplicial geometry defined by the $J_{b}$ values
as before. This means in the asymptotic limit the foliating simplicial
3-surfaces act as space-like simplicial $3$-surfaces of a simplicial
4-geometry defined by the $J_{b}$ values. This suggests that in the asymptotic
limit a certain sense of time exists in the order of the foliating hypersurfaces.

In case of a $H^{-}\approx SL(2,C)/SU(1,1)$ based spin foam model the $J_{b}$
are both discrete and continuous \cite{R4}. The spin network functionals for
the Lorentzian quantum gravity are being currently studied and will be
published elsewhere.

\section{Discussion and comments.\label{sec.3}}

Now let us compare our formalism in the previous section to that of the
canonical quantum formulation.

\begin{itemize}
\item The Gauss constraint has been implemented in our formalism by the use of
the gauge invariant spin network functionals for the quantum states. There is
an important difference between the two formulations in the case of the
Lorentzian quantum gravity. It is that the spin network functionals are made
of the finite dimensional representations \cite{RS} in the canonical
formalism, while here they are made of infinite dimensional representations
\cite{bc2}. This difference needs to be investigated.

\item The coordinate independence has been implemented here at the classical
level by the use of the discretized action.

\item The Hamiltonian constraint of canonical quantum gravity contains
evolution information. So it essentially should be contained in the definition
of the elementary transition amplitudes given in equation (\ref{eq.gr.tr}).
\end{itemize}

Our formulation has brought the spin foams closer to canonical quantum gravity
in the formal set up and in certain details. Our formulation has both the
features of spin foam models and the canonical formulation. Since the spin
foams are derived from the discretized action, it is reasonable to say that
the canonical formulation can be further related to the spin foam model of
gravity by studying the continuum limit. But before that, we believe the
imposition of Barrett-Crane constraints on the BF spin foams has to be
rigorously derived from a discrete action and the amplitudes of the lower
dimensional simplices fixed (please see \cite{AP3} for more discussion on this).

One of the problems with canonical quantum gravity is in defining a proper
Hamiltonian constraint operator. The proposal by Thiemann \cite{TT1} for a
Hamiltonian constraint operator appears to be set back by anomalies \cite{P1}.
By studying the continuum limit of our elementary transition amplitudes one
might be able to get a useful physical Hamiltonian operator (physical inner
product) for canonical quantum gravity.

There are many open questions that need to be addressed, such as:

\begin{itemize}
\item What can we learn from this approach about the physics of quantum
gravity? For example, is quantum gravity unitary?

\item What are the potential applications to the physical problems?

\item What is the continuum limit?

\item How to include the topologies in our theory that were excluded by the
conditions that were specified in the beginning of section three?

\item How to include matter?
\end{itemize}

\section{Acknowledgements.}

I am grateful to George Sparling for discussions, encouragement and guidance
in developing this article and in learning the material. I thank Allen Janis
for discussions, support and encouragement. I thank Jorge Pullin and Alejandro
Perez for discussions and correspondences while learning the foundations for
this research. I\ thank John Baez for correspondences regarding this article
and in helping me learn the material.

\appendix{}

\section{Calculation of edge integrals for compact groups.}

Let $G$ be a compact group. Intertwiners are required for the calculation of
the following integral, which we refer to as the edge integral:%
\begin{equation}
\int dg\bigotimes_{l}\rho_{J_{l}}(g)=\sum_{i}i\bar{\imath}. \label{eq.i}%
\end{equation}
where the bar denotes adjoint operation.

Explicitly, the above equation is
\begin{equation}
\int dg\rho_{n_{1}J_{b_{1}}}^{m_{1}}(g)\rho_{n_{2}J_{b_{1}}}^{m_{2}}%
(g)...\rho_{n_{3}J_{b_{3}}}^{m_{3}}(g)=\sum_{X}i_{J_{b_{1}}J_{b_{2}%
}...J_{b_{N}}X}^{m_{1}m_{2}...m_{N}}\bar{\imath}_{n_{1}n_{2}...n_{N}%
}^{J_{b_{1}}J_{b_{2}}...J_{b_{N}}X}, \label{eq:1}%
\end{equation}
where $b_{1},b_{2}...b_{N}$ are the bones that pass through an edge $e$. Each
value of $X$ identifies a unique intertwiner.

\begin{figure}
[ptbh]
\begin{center}
\includegraphics[
height=3.3122in,
width=2.4405in
]%
{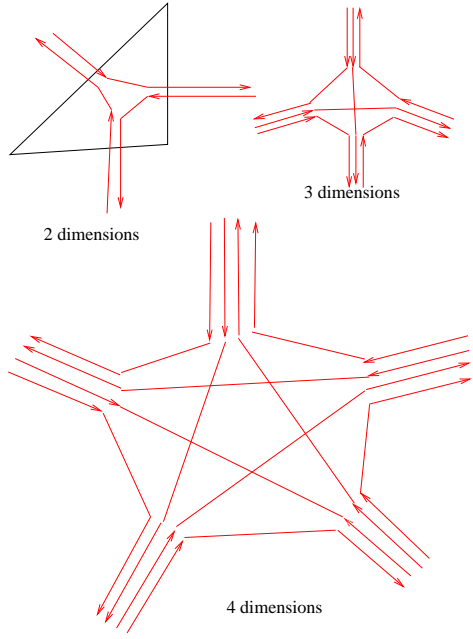}%
\caption{Holonomy Directions.}%
\label{fig.holdir}%
\end{center}
\end{figure}

In the calculation of the above edge integral, it is assumed that the
holonomies are traversed through the edge of a simplex in the same direction
as in the derivation of the BF spin foam model in section two. But usually the
directions are random. Reversing the direction of a holonomy is equivalent to
complex conjugating (the inverse of the transpose) the representations in the
edge integral. To simplify the calculation of the edge integrals, the
directions of the holonomies can be chosen appropriately, as illustrated in
figure$\ $(\ref{fig.holdir}) in $2$, $3$ and $4$ dimensions.%

For convenience we adjoint one or more of the $\rho$'s as needed which is
equivalent to choosing the direction of the holonomies. Let $\alpha_{J_{1}%
}^{m_{1}}$, $\beta_{J_{2}}^{m_{2}}$ be the basis of the $G$-vector components
in the $J_{1}$ and $J_{2}$ representations. Then the tensor product of these
two can be expanded as follows in terms of the Clebsch-Gordan coefficients:%
\begin{equation}
\alpha_{J_{1}}^{m_{1}}\beta_{J_{2}}^{m_{2}}=\sum_{\{J_{3},r\}}C_{\{J_{3}%
,r\}m_{3}}^{J_{1}m_{1}J_{2}m_{2}}\gamma_{\{J_{3},r\}}^{m_{3}},
\end{equation}
where the $\gamma_{J_{3}r}^{m_{3}}$ are the components of a $G$-vector in the
$J_{3}$ representation. The variable $r$ denotes the various copies of the
same representation in the outer sum.

Let $d_{J_{b}}$ be the dimension of the $J_{b}$ representation of the group.
The intertwiners are calculated using the following two identities:%
\begin{equation}
\int dg\rho_{n_{1}J_{1}}^{m_{1}}(g)\overline{\rho}_{n_{2}J_{2}}^{m_{2}%
}(g)=\frac{1}{d_{J_{1}}}\delta_{J_{1}J_{1}}\delta^{m_{1}m_{2}}\delta
_{n_{1}n_{2}}, \label{eq.I.1}%
\end{equation}

and
\begin{equation}
\rho_{n_{1}J_{1}}^{m_{1}}(g)\rho_{n_{2}J_{2}}^{m_{2}}(g)={\sum_{J_{3}r}}%
\sum_{m_{3},n_{3}}C_{\{J_{3},r\}m_{3}}^{J_{1}m_{1}J_{2}m_{2}}C_{J_{1}%
n_{1}J_{2}n_{2}}^{\{J_{3},r\}n_{3}}\rho_{_{n_{3}}J_{3}}^{m_{3}}(g),
\label{eq:I.2}%
\end{equation}
where the $C_{\{J_{3},r\}m_{3}}^{J_{1}m_{1}J_{2}m_{2}}$ are the Clebsch-Gordan
coefficients, $C_{J_{1}n_{1}J_{2}n_{2}}^{\{J_{3},r\}n_{3}}$ is the adjoint of
$C_{\{J_{3},r\}n_{3}}^{J_{1}n_{1}J_{2}n_{2}}$. $C_{\{J_{3},r\}n_{3}}%
^{J_{1}n_{1}J_{2}n_{2}}$ is also the inverse of $C_{J_{1}n_{1}J_{2}n_{2}%
}^{\{J_{3},r\}n_{3}}$ because of unitarity. I refer to \cite{de4} for more information.

From equation (\ref{eq.I.1}) we can define the intertwiners in two dimensional
space:%
\begin{equation}
i_{J_{b_{1}}\bar{J}_{b_{2}}}^{m_{b_{1}}m_{b_{2}}}=\bar{\imath}_{m_{b_{1}%
}m_{b_{2}}}^{J_{b_{1}}\bar{J}_{b_{2}}}=\frac{1}{\sqrt{d_{J_{b1}}}}%
\delta_{J_{b_{1}}J_{b_{2}}}\delta^{m_{b_{1}}m_{b_{2}}}. \label{eq.2int}%
\end{equation}
where the $\bar{J}$ is the conjugate representation of $J.$

\begin{figure}
[h]
\begin{center}
\includegraphics[
height=1.3206in,
width=4.4849in
]%
{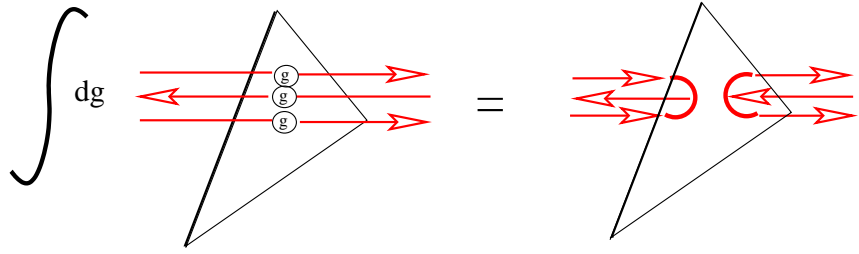}%
\caption{The edge integral in three dimensions.}%
\label{intertwine.fig}%
\end{center}
\end{figure}

The edge integral in equation (\ref{eq:1}) in three dimensions, using
equations (\ref{eq.I.1}) and (\ref{eq:I.2}), is given by%
\[
\int dg\rho_{n_{1}J_{1}}^{m_{1}}(g)\rho_{n_{2}J_{2}}^{m_{2}}(g)\bar{\rho
}_{n_{3}J_{3}}^{m_{3}}(g)=\int dg{\sum_{J_{,}t}}C_{\{J,t\}m}^{J_{1}m_{1}%
J_{2}m_{2}}C_{J_{1}n_{1}J_{2}n_{2}}^{\{J,t\}n}\rho_{nJ}^{m}(g)\bar{\rho
}_{n_{3}J_{3}}^{m_{3}}(g)
\]%
\begin{equation}
={\sum_{J_{,}t}}\frac{1}{d_{J_{3}}}C_{\{J_{3},t\}m}^{J_{1}m_{1}J_{2}m_{2}%
}C_{J_{1}n_{1}J_{2}n_{2}}^{\{J_{3},t\}n} \label{intertwine.eq}%
\end{equation}
This calculation has summarised in figure (\ref{intertwine.fig}). From
equation (\ref{intertwine.eq}) the intertwiners can be defined by comparing
this to (\ref{eq.i}) as follows:%
\begin{equation}
i_{J_{1}J_{2}\bar{J}_{3},t}^{m_{1}m_{2}m}=\frac{1}{\sqrt{d_{J_{3}}}}%
C_{\{J_{3},t\}m}^{J_{1}m_{1}J_{2}m_{2}}.
\end{equation}
Also we can identify that the role of variable $X$ in (\ref{eq:1}) is played
here by $t$. The calculation of 2D edge integrals is summarised in figure
(\ref{intertwine.fig}), where the summation over t is not shown.%

\section{ Edge integrals in four dimensions.}

\begin{figure}
[ptbh]
\begin{center}
\includegraphics[
height=1.7443in,
width=1.983in
]%
{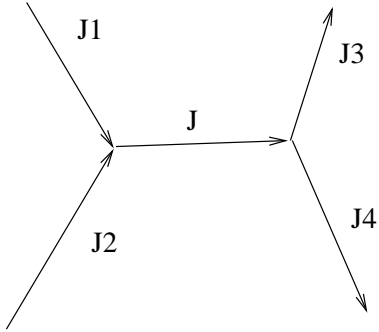}%
\caption{4D $SO(4)$ BF Intertwiner}%
\label{4J.fig}%
\end{center}
\end{figure}

Here we calculate the following edge integral, which is written according to
the directions for the holonomies in figure (\ref{fig.holdir}):%
\[
\int dg\rho_{n_{1}J_{1}}^{m_{1}}(g)\rho_{n_{2}J_{2}}^{m_{2}}(g)\bar{\rho
}_{n_{3}J_{3}}^{m_{3}}(g)\bar{\rho}_{n_{4}J_{4}}^{m_{4}}(g)
\]%
\[
=\int dg{\sum_{J_{5,}t}}C_{\{J_{5},t\}m_{5}}^{J_{1}m_{1}J_{2}m_{2}}%
C_{J_{1}n_{1}J_{2}n_{2}}^{\{J_{5},t\}n_{5}}\rho_{_{n_{5}}J_{5}}^{m_{5}%
}(g){\sum_{J_{6,}r}}C_{J_{3}m_{3}J_{4}m_{4}}^{\{J_{6},r\}m_{6}r}%
C_{\{J_{6},r\}n_{6}}^{J_{3}n_{3}J_{4}n_{4}}\bar{\rho}_{_{n_{6}}J_{6}}^{m_{6}%
}(g)
\]%
\[
={\sum_{J,r,t}}\sum_{k,l}\frac{1}{d_{J}}C_{\{J,t\}k}^{J_{1}m_{1}J_{2}m_{2}%
}C_{J_{1}n_{1}J_{2}n_{2}}^{\{J,t\}l}C_{J_{3}m_{3}J_{4}m_{4}}^{\{J,r\}k}%
C_{\{J,r\}l}^{J_{3}n_{3}J_{4}n_{4}},
\]
from which we can identify the intertwiners and $X$ as%
\begin{align*}
i_{J_{1}J_{2}\bar{J}_{3}\bar{J}_{4}\{J,t,r\}}^{m_{1}m_{2}m_{3}m_{4}}  &
=\frac{1}{\sqrt{d_{J}}}\sum_{l}C_{J_{1}m_{1}J_{2}m_{2}}^{\{J,r\}l}%
C_{\{J,t\}l}^{J_{3}m_{3}J_{4}m_{4}}\\
X  &  =\{J,t,r\},
\end{align*}
which can be diagrammatically represented as in figure (\ref{4J.fig}).%

\section{The Barrett-Crane intertwiner.}

Riemannian quantum gravity is built on the representation theory of $SO(4).$
Because of the isomorphism $SO(4)$ $\cong$ $SU(2)\otimes SU(2),$ each
irreducible representation$\ $of $SO(4)$ is labelled by a pair of $SU(2)$
representations $(J_{L,}J_{R}).$ The Clebsch-Gordan coefficients of $SO(4)$
are just the tensor product of two $SU(2)$ Clebsch-Gordan coefficients. Since
the $SU(2)$ representations are conjugate equivalent, so are the
representations of $SO(4)$. Application of the Barrett-Crane constraints
restricts the representations to those for which $J_{L}=J_{R}$ \cite{bc1}.
These are called the simple representations. The Barrett-Crane intertwiner is
defined using the Clebsch-Gordan coefficients as given below, where the $C$
are the Clebsch-Gordan coefficient for $SO(4)$ (no multiplicities), with all
the $J$'s restricted to simple representations:%
\begin{equation}
i_{m_{1}m_{2}m_{3}m_{4}}^{J_{1}J_{2}\bar{J}_{3}\bar{J}_{4}}=\sum_{J}\frac
{1}{d_{J}}\sum_{k}C_{J_{1}m_{1}J_{2}m_{2}}^{Jk}C_{{}Jk}^{J_{3}m_{3}J_{4}m_{4}%
}. \label{eq.bc}%
\end{equation}

An important property of the above intertwiner is that, it does not depend on
how you make the split in the four $J$'s into two pair of $J$'s, to write the
right hand side.

The above intertwiner can be written in a different way. Each $m_{i}$ in
equation (\ref{eq.bc}) can be explicitly represented as a pair, $\left(
l_{i},r_{i}\right)  .$ So equation (\ref{eq.bc}) can be rewritten as follows:%
\[
i_{l_{1}r_{1}l_{2}r_{2}l_{3}r_{3}l_{4}r_{4}}^{J_{1}J_{2}\bar{J}_{3}\bar{J}%
_{4}}=\sum_{J}\frac{1}{d_{J}}\sum_{l,r}C_{J_{1}r_{1}J_{2}r_{2}}^{Jr}%
C_{Jl}^{J_{1}l_{1}J_{2}l_{2}}C_{J_{3}l_{3}J_{4}{}l_{4}}^{Jl}C_{{}Jr}%
^{J_{3}r_{3}J_{4}r_{4}}%
\]%
\[
=\sum_{J}d_{J}\int dh\rho_{r_{1}J_{1}}^{l_{1}}(h)\rho_{r_{2}J_{2}}^{l_{2}%
}(h)\bar{\rho}_{rJ}^{l}(h)\int dh\bar{\rho}_{r_{3}J_{3}}^{l_{3}}(\grave
{h})\bar{\rho}_{r_{4}J_{4}}^{l_{4}}(\grave{h})\rho_{rJ}^{l}(\grave{h})
\]%
\[
=\int dh\rho_{r_{1}J_{1}}^{l_{1}}(h)\rho_{r_{2}J_{2}}^{l_{2}}(h)\int
d\grave{h}\bar{\rho}_{r_{3}J_{3}}^{l_{3}}(\grave{h})\bar{\rho}_{r_{4}J_{4}%
}^{l_{4}}(\grave{h})\sum_{J}d_{J}\bar{\rho}_{r_{3}J}^{l_{3}}(h)\rho_{rJ}%
^{l}(\grave{h})
\]%
\[
=\int dh\rho_{r_{1}J_{1}}^{l_{1}}(h)\rho_{r_{2}J_{2}}^{l_{2}}(h)\int
d\grave{h}\bar{\rho}_{r_{3}J_{3}}^{l_{3}}(\grave{h})\bar{\rho}_{r_{4}J_{4}%
}^{l_{4}}(\grave{h})\delta(h^{-1}\grave{h})
\]%
\[
=\int dh\rho_{r_{1}J_{1}}^{l_{1}}(h)\rho_{r_{2}J_{2}}^{l_{2}}(h)\bar{\rho
}_{r_{3}J_{3}}^{l_{3}}(h)\bar{\rho}_{r_{4}J_{4}}^{l_{4}}(\grave{h})
\]%
\begin{equation}
=\int dh\rho_{r_{1}J_{1}}^{l_{1}}(h)\rho_{r_{2}J_{2}}^{l_{2}}(h)\rho
_{l_{3}J_{3}}^{r_{3}}(h^{-1})\rho_{l_{4}J_{4}}^{r_{4}}(h^{-1}),
\end{equation}
where $\grave{h}$ and $h$ belong to $SU(2)$.

Restricting the representation to simple ones effectively reduces the harmonic
analysis on $SO(4)$ to $S^{3}.$

In the last equation, $h$ must be seen as an element of $S^{3}$ instead of
$SU(2)$. Let $h:S^{3}\rightarrow SU(2)$ is a bijective mapping. Then the
Barrett-Crane intertwiner can be rewritten as%
\begin{equation}
i_{l_{1}r_{1}l_{2}r_{2}l_{3}r_{3}l_{4}r_{4}}^{J_{1}J_{2}\bar{J}_{3}\bar{J}%
_{4}}=\int_{S^{3}}dx\rho_{r_{1}J_{1}}^{l_{1}}(h(x))\rho_{r_{2}J_{2}}^{l_{2}%
}(h(x))\rho_{l_{3}J_{3}}^{r_{3}}(h^{-1}(x))\rho_{l_{4}J_{4}}^{r_{4}}%
(h^{-1}(x)). \label{eq.bc1}%
\end{equation}


\begin{thebibliography}{99}                                                                                               %


\bibitem {BF}A.S. Schwarz, Lett. Math. Phys. 2 (1978) 247; Commun. Math. Phys.
67 (1979) 1; G. Horowitz, Commun. Math. Phys. 125 (1989) 417; M. Blau and G.
Thompson, Ann. Phys. 209 (1991) 129.

\bibitem {Rg1}G. Ponzano and T. Regge, Semiclassical Limit of Racah
Coefficients, Spectroscopy and Group Theoretical Methods in Physics, F. Block
et al (Eds), North-Holland, Amsterdam, 1968.

\bibitem {RP}R. Penrose-Angular momentum: an approach to combinatorial
space-time, Quantum Theory and Beyond, essays and discussions arising from a
colloquium, Edited by Ted Bastin, Cambridge University Press, 1971.

\bibitem {R5}C. Rovelli, The projector on physical states in loop quantum
gravity, Preprint available as arXiv:gr-qc/9806121; C. Rovelli and M. P.
Reisenberger, "Sum over Surfaces" form of Loop Quantum Gravity, arXiv:gr-qc/9612035.

\bibitem {fms}F. Markopoulou and L. Smolin, Causal evolution of spin networks,
Nucl.Phys. B,508:409--430, 1997.

\bibitem {fm}F. Markopoulou, Dual formulation of spin network evolution,
Preprint available as gr-qc/9704013.

\bibitem {bz1}J. C. Baez, An Introduction to Spin Foam Models of Quantum
Gravity and BF Theory, Lect.Notes.Phys., 543:25--94, 2000

\bibitem {AP2}A. Perez, Spin Foam Models for Quantum Gravity, Preprint
available as arXiv:gr-qc/0301113.

\bibitem {oo2}H. Ooguri, Topological Lattice Models in Four Dimensions,
Mod.Phys.Lett. A, 7:2799--2810, 1992.

\bibitem {bc1}J. W. Barrett and L. Crane, Relativistic Spin Networks and
Quantum Gravity. J.Math.Phys., 39:3296--3302, 1998.

\bibitem {As}A. Ashtekar, Lectures on non Perturbative Canonical Gravity, Word
Scientific, 1991.

\bibitem {RS}C. Rovelli and L. Smolin, Spin Networks and Quantum Gravity,
Preprint available as gr-qc 9505006.

\bibitem {TT1}T. Thiemann, Anomaly-free formulation of non-perturbative,
four-dimensional Lorentzian quantum gravity, Preprint available as arXiv:gr-qc/9606088.

\bibitem {R3}R. Borissov, R. Pietri and C. Rovelli, Matrix Elements of
Thiemann's Hamiltonian Constraint in Loop Quantum Gravity, Preprint available
as arXiv:gr-qc/9703090.

\bibitem {rz2}M. P. Reisenberger, On Relativistic Spin Network Vertices, J.
Math. Phys., 40:2046--2054, 1999.

\bibitem {Lfr1}R. De. Pietri and L. Freidel, SO(4) Plebanski Action and
Relativistic Spin Foam Model. Class.Quant.Grav., 16:2187--2196, 1999.

\bibitem {de1}N. Ja. Vilenkin and A. U. Klimyk, Representation theory of Lie
groups and Special Functions. Volume 1, Kluwer Academic Publisher, Dordrecht,
The Netherlands, 1993.

\bibitem {AP}A. Perez, Spin Foam Models for Quantum Gravity, Ph.D. Thesis,
University of Pittsburgh, Pittsburgh, USA 2001.

\bibitem {Barb}J. F. Barbero, Real Ashtekar Variables for Lorentzian Signature
Space Times. Phys. Rev., D51:5507--5510, 1995.

\bibitem {rz4}M. P. Reisenberger, Worldsheet formulations of gauge theories
and gravity", gr-qc/9412035

\bibitem {Rg2}T. Regge, General Relativity without Coordinates, Nuovo Cimento
19 (1961) 558-571.

\bibitem {bc2}J. W. Barrett and L. Crane, A Lorentzian Signature Model for
Quantum General Relativity, Class.Quant.Grav., 17:3101--3118, 2000.

\bibitem {BS}J. W. Barrett and C. Steele, Asymptotics of relativistic spin
networks, Preprint available as arXiv:gr-qc/0209023..

\bibitem {R4}A. Perez and C. Rovelli, 3+1 Spin foam Model of Quantum gravity
with Spacelike and Timelike Components, Preprint available as arXiv:gr-qc/0011037.

\bibitem {AP3}M. Bojowald and A. Perez, Spin Foam Quantization and Anomalies,
Preprint gr--qc/0303026

\bibitem {P1}R. Gambini, J. Lewandowski, D. Marolf and G. Pullin, On the
consistency of the constraint algebra in spin network quantum gravity,
Preprint available as arXiv:gr-qc/9710018

\bibitem {de4}N. Ja. Vilenkin and A. U. Klimyk Representation theory of Lie
groups and Special Functions, Recent advances, Kluwer Academic Publisher,
Dordrecht, The Netherlands, 1995.

\bibitem {bz2}J. C. Baez, Spin Foam Models, Class Quant Grav 15 (1998)
1827-1858; gr-qc/9709052.

\bibitem {bz3}J. C. Baez, J. D. Christenson, Thomas R. Halford, and D. C.
Tsang, Spin foam models of Riemannian quantum gravity, Preprint arXiv:gr-qc/0202017.

\bibitem {R1}C. Rovelli, The basis of the Ponzano-Regge-Turaev-Viro-Ooguri
quantum-gravity model is the Loop Representation basis, Preprint available as arXiv:hep-th/9304164.

\bibitem {or1}D. Oriti and R. M.Williams, Gluing 4-simplices: a derivation of
the Barrett-Crane spin foam model for Euclidean quantum gravity, Preprint
available: arXiv:gr-qc/0010031.

\bibitem {vmk}D. A. Varshalovich, A. N. Moskalev and V. K. Khersonskii,
Quantum Theory of Angular Momentum, World Scientific, 1988.

\bibitem {TV}V. Turaev and O. Viro, State sum invariants of 3-manifolds and
quantum 6j symbols, Topology 31(1992), 865-902.

\bibitem {rz3}M. P. Reisenberger, Worldsheet formulations of gauge theories
and gravity, arXiv:gr-qc/9412035.

\bibitem {R6}A. Perez, C. Rovelli, Observables in quantum gravity, Preprint arXiv:gr-qc/0104034.

\bibitem {MA}M. Arnsdorf, Relating canonical and covariant approaches to
triangulated models of quantum gravity, Preprint arXiv:gr-qc/0110026.

\bibitem {EL}E. R. Livine, Projected Spin Networks for Lorentz connection:
Linking Spin Foams and Loop Gravity, Preprint arXiv:gr-qc/0207084.

\bibitem {R7}A. Perez and C. Rovelli, Spin foam model for Lorentzian General
Relativity, Preprint arXiv:gr-qc/0009021

\bibitem {Lfr}L. Freidel and K. Krasnov, Simple Spin Networks as Feynman
Graphs, Preprint available as hep-th/9903192.

\bibitem {oo1}H. Ooguri, Partition Functions and Topology-Changing Amplitudes
in the 3D Lattice Gravity of Ponzano and Regge, Preprint available as arXiv:hep-th/9112072.

\bibitem {rz1}M. P. Reisenberger, Classical Euclidean general relativity from
"left-handed area = right-handed area", arXiv:gr-qc/9804061.
\end{thebibliography}
\end{document}